\documentclass[10pt]{article}
\usepackage{pst-all}
\usepackage{letterspace}
\usepackage{psfrag}
\usepackage{latexsym}
\usepackage{amsmath}
\usepackage{amssymb}
\usepackage[dvips]{graphicx}
\usepackage{caption2}
\usepackage[noend]{algorithmic}
\usepackage[plain]{algorithm}

\usepackage{color}
\usepackage[nohead, top=2.8cm, bottom=1.6cm, left=2.9cm,
right=3.1cm]{geometry}

\def\yskip{\penalty-50\vskip3pt plus3pt minus2pt}
\def\y{\yskip}
\def\yy{\yskip\yskip}
\def\yyy{\yskip\yskip\yskip}
\def\qed{\hskip 3pt\vrule height 6pt width 3pt depth 0pt}
\def\q5uad{\quad\quad\quad\quad\quad}
\def\mytab{\phantom{xxx}}

\def\s{\ }

\title{\bf Linear Coloring and Linear Graphs\,\vspace{0.4cm}\footnote{This research is co-financed by E.U.-European Social
Fund (75\%) and the Greek Ministry of Development-GSRT (25\%). }}

\author{\large Kyriaki Ioannidou \ and \ Stavros D. Nikolopoulos}

\date{}

\begin{document}

\maketitle

\vspace{-0.6cm}

\centerline{\it Department of Computer Science, University of
Ioannina}

\centerline{\it P.O.Box 1186, GR-45110 \s Ioannina, Greece}

\centerline{\tt \{kioannid, stavros\}@cs.uoi.gr}

\vskip 0.3in

\begin{center}
\noindent
\parbox{5.5in}
{{\bf Abstract:} \s Motivated by the definition of linear coloring
on simplicial complexes, recently introduced in the context of
algebraic topology \cite{Civan}, and the framework through which
it was studied, we introduce the linear coloring on graphs. We
provide an upper bound for the chromatic number $\chi(G)$, for any
graph $G$, and show that $G$ can be linearly colored in polynomial
time by proposing a simple linear coloring algorithm. Based on
these results, we define a new class of perfect graphs, which we
call co-linear graphs, and study their complement graphs, namely
linear graphs. The linear coloring of a graph $G$ is a vertex
coloring such that two vertices can be assigned the same color, if
their corresponding clique sets are associated by the set
inclusion relation (a clique set of a vertex $u$ is the set of all
maximal cliques containing $u$); the linear chromatic number
$\mathcal{\lambda}(G)$ of $G$ is the least integer $k$ for which
$G$ admits a linear coloring with $k$ colors. We show that linear
graphs are those graphs $G$ for which the linear chromatic number
achieves its theoretical lower bound in every induced subgraph of
$G$. We prove inclusion relations between these two classes of
graphs and other subclasses of chordal and co-chordal graphs, and
also study the structure of the forbidden induced subgraphs of the
class of linear graphs.


}
\end{center}

\vskip 0.3in 
\section{Introduction}

\noindent {\bf Framework-Motivation.} A {\it linear coloring} of a
graph $G$ is a coloring of its vertices such that if two vertices
are assigned the same color, then their corresponding clique sets
are associated by the set inclusion relation; a {\it clique set}
of a vertex $u$ is the set of all maximal cliques in $G$
containing $u$. The linear chromatic number $\mathcal{\lambda}(G)$
of $G$ is the least integer $k$ for which $G$ admits a linear
coloring with $k$ colors.

\y Motivated by the definition of linear coloring on simplicial
complexes associated to graphs, first introduced by Civan and
Yal\c{c}in \cite{Civan} in the context of algebraic topology, we
define the linear coloring on graphs. The idea for translating
their definition in graph theoretic terms came from studying
linear colorings on simplicial complexes which can be represented
by a graph. In particular, we studied the linear coloring on the
independence complex $\mathcal{I}(G)$ of a graph $G$, which can
always be represented by a graph and, more specifically, is
identical to the complement graph $\skew3\overline{G}$ of $G$ in
graph theoretic terms; indeed, the facets of $\mathcal{I}(G)$ are
exactly the maximal cliques of $\skew3\overline{G}$. However, the
two definitions cannot always be considered as identical since not
in all cases a simplicial complex can be represented by a graph;
such an example is the neighborhood complex $\mathcal{N}(G)$ of a
graph $G$. Recently, Civan and Yal\c{c}in \cite{Civan} studied the
linear coloring of the neighborhood complex $\mathcal{N}(G)$ of a
graph $G$ and proved that, for any graph $G$, the linear chromatic
number of $\mathcal{N}(G)$ gives an upper bound for the chromatic
number of the graph $G$. This approach lies in a general framework
met in algebraic topology.

\y In the context of algebraic topology, one can find much work
done on providing boundaries for the chromatic number of an
arbitrary graph $G$, by examining the topology of the graph
through different simplicial complexes associated to the graph.
This domain was motivated by Kneser's conjecture, which was posed
in 1955, claiming that ``if we split the $n$-subsets of a
$(2n+k)$-element set into $k+1$ classes, one of the classes will
contain two disjoint $n$-subsets" \cite{Kne55}. Kneser's
conjecture was first proved by Lov\'{a}sz in 1978, with a proof
based on graph theory, by rephrasing the conjecture into ``the
chromatic number of Kneser's graph $KG_{n,k}$ is $k+2$"
\cite{Lov1978}. Many more topological and combinatorial proofs
followed the interest of which extends beyond the original
conjecture \cite{Zieg02}. Although Kneser's conjecture is
concerned with the chromatic numbers of certain graphs (Kneser
graphs), the proof methods that are known provide lower bounds for
the chromatic number of any graph \cite{Mat04}. Thus, this
initiated the application of topological tools in studying graph
theory problems and more particularly in graph coloring problems
\cite{Csorba04}.

\y The interest to provide boundaries for the chromatic number
$\chi(G)$ of an arbitrary graph $G$ through the study of different
simplicial complexes associated to $G$, which is found in
algebraic topology bibliography, drove the motivation for defining
the linear coloring on the graph $G$ and studying the relation
between the chromatic number $\chi(G)$ and the linear chromatic
number $\lambda(\skew3\overline{G})$. We show that for any graph
$G$, $\lambda(\skew3\overline{G})$ is an upper bound for
$\chi(G)$. The interest of this result lies on the fact that we
present a linear coloring algorithm that can be applied to any
graph $G$ and provides an upper bound
$\lambda(\skew3\overline{G})$ for the chromatic number of the
graph $G$, i.e. $\chi(G) \leq \lambda(\skew3\overline{G})$; in
particular, it provides a proper vertex coloring of $G$ using
$\lambda(\skew3\overline{G})$ colors. Additionally, recall that a
known lower bound for the chromatic number of any graph $G$ is the
clique number $\omega(G)$ of $G$, i.e. $\chi(G) \geq \omega(G)$.
Motivated by the definition of perfect graphs, for which
$\chi(G_A)=\omega(G_A)$ holds $\forall A \subseteq V(G)$, it was
interesting to study those graphs for which the equality $\chi(G)
= \lambda(\skew3\overline{G})$ holds, and even more those graphs
for which this equality holds for every induced subgraph. The
outcome of this study was the definition of a new class of perfect
graphs, namely co-linear graphs, and, furthermore, the study of
the classes of co-linear graphs and of their complement class,
namely linear graphs.

\yy \noindent {\bf Our Results.} In this paper, we first introduce
the linear coloring of a graph~$G$ and study the relation between
the linear coloring of~$\skew3\overline{G}$ and the proper vertex
coloring of~$G$. We prove that, for any graph $G$, a linear
coloring of $\skew3\overline{G}$ is a proper vertex coloring of
$G$ and, thus, $\lambda(\skew3\overline{G})$ is an upper bound for
$\chi(G)$, i.e. $\chi(G) \leq \lambda(\skew3\overline{G})$. We
present a linear coloring algorithm that can be applied to any
graph $G$. Motivated by these results and the Perfect Graph
Theorem \cite{Gol}, we study those graphs for which the equality
$\chi(G) = \lambda(\skew3\overline{G})$ holds for every induce
subgraph and define a new class of perfect graphs, namely
co-linear graphs; we also study their complement class, namely
linear graphs. A graph~$G$ is a {\it co-linear graph} if and only
if its chromatic number $\mathcal{\chi}(G)$ equals to the linear
chromatic number $\mathcal{\lambda}(\skew3\overline{G})$ of its
complement graph $\skew3\overline{G}$, and the equality holds for
every induced subgraph of $G$, i.e.
$\mathcal{\chi}(G_A)=\mathcal{\lambda}(\skew3\overline{G}_A)$,
$\forall A \subseteq V(G)$; a graph $G$ is a {\it linear graph} if
it is the complement of a co-linear graph. We show that the class
of co-linear graphs is a superclass of the class of threshold
graphs, a subclass of the class of co-chordal graphs and is
distinguished from the class of split graphs. Additionally, we
give some structural and recognition properties for the classes of
linear and co-linear graphs. We study the structure of the
forbidden induced subgraphs of the class of linear graphs, and
show that any $P_6$-free chordal graph, which is not a linear
graph, properly contains a $k$-sun as an induced subgraph.
Therefore, we infer that the subclass of chordal graphs, namely
linear graphs, is a superclass of the class of $P_6$-free strongly
chordal graphs.

\yy \noindent {\bf Basic Definitions.} Some basic graph theory
definitions follow. We consider finite undirected and directed
graphs with no loops or multiple edges. Let $G$ be such a graph;
then, $V(G)$ and $E(G)$ denote the set of vertices and of edges of
$G$, respectively. An edge is a pair of distinct vertices $x, y
\in V(G)$, and is denoted by $xy$ if $G$ is an undirected graph
and by $\overrightarrow{xy}$ if $G$ is a directed graph. For a set
$A \subseteq V(G)$ of vertices of the graph $G$, the subgraph of
$G$ {\it induced\/} by $A$ is denoted by $G_A$. Additionally, the
cardinality of a set $A$ is denoted by $|A|$. For a given vertex
ordering $(v_1,v_2,\ldots,v_n)$ of a graph $G$, the subgraph of
$G$ induced by the set of vertices $\{v_i,v_{i+1},\ldots,v_n\}$ is
denoted by $G_i$. The set $N(v)=\{u \in V(G) : (u,v) \in E(G)\}$
is called the {\it open neighborhood} of the vertex $v \in V(G)$
in $G$, sometimes denoted by $N_G(v)$ for clarity reasons. The set
$N[v]=N(v) \cup \{v\}$ is called the {\it closed neighborhood} of
the vertex $v \in V(G)$ in $G$. In a graph $G$, the length of a
path is the number of edges in the path. The {\it distance}
$d(v,u)$ from vertex $v$ to vertex $u$ is the minimum length of a
path from $v$ to $u$; $d(v,u)=\infty$ if there is no path from $v$
to $u$.

\y The greatest integer $r$ for which a graph $G$ contains an
independent set of size $r$ is called the {\it independence
number} or otherwise the {\it stability number} of $G$ and is
denoted by $\mathcal{\alpha}(G)$. The cardinality of the vertex
set of the maximum clique in $G$ is called the {\it clique number}
of $G$ and is denoted by $\mathcal{\omega}(G)$. A {\it proper
vertex coloring} of a graph $G$ is a coloring of its vertices such
that no two adjacent vertices are assigned the same color. The
{\it chromatic number} $\mathcal{\chi}(G)$ of $G$ is the least
integer $k$ for which $G$ admits a proper vertex coloring with $k$
colors. For the numbers $\mathcal{\omega}(G)$ and
$\mathcal{\chi}(G)$ of an arbitrary graph $G$ the inequality
$\mathcal{\omega}(G) \leq \mathcal{\chi}(G)$ holds. In
particularly, $G$ is a {\it perfect graph} if the equality
$\mathcal{\omega}(G_A) = \mathcal{\chi}(G_A)$ holds $\forall A
\subseteq V(G)$. For more details on basic definitions in graph
theory refer to \cite{Brand99, Gol}.

\y Next, definitions of some graph classes mentioned throughout
the paper follow. A graph is called a {\it chordal graph} if it
does not contain an induced subgraph isomorphic to a chordless
cycle of four or more vertices. A graph is called a {\it
co-chordal graph} if it is the complement of a chordal graph
\cite{Gol}. A hole is a chordless cycle $C_n$ if $n \ge 5$; the
complement of a hole is an antihole. A graph $G$ is a {\it split
graph} if there is a partition of the vertex set $V(G)=K+I$, where
$K$ induces a clique in $G$ and $I$ induces an independent set.
Split graphs are characterized as $(2K_2,C_4,C_5)$-free. {\it
Threshold graphs} are defined as those graphs where stable subsets
of their vertex sets can be distinguished by using a single linear
inequality. Threshold graphs were introduced by Chv\'{a}tal and
Hammer \cite{Chv77} and characterized as $(2K_2,P_4,C_4)$-free.
{\it Quasi-threshold} graphs are characterized as the
$(P_4,C_4)$-free graphs and are also known in the literature as
trivially perfect graphs \cite{Gol, Nik00}. A graph is {\it
strongly chordal} if it admits a strong perfect elimination
ordering. Strongly chordal graphs were introduced by Farber in
\cite{Far83} and are characterized completely as those chordal
graphs which contain no $k$-sun as an induced subgraph. For more
details on basic definitions in graph theory refer to
\cite{Brand99, Gol}.

\section{Linear Coloring on Graphs}

In this section we define the linear coloring of a graph $G$, we
prove some properties of the linear coloring of $G$, and present a
simple algorithm for linear coloring that can be applied to any
graph $G$. It is worth noting that similar properties of linear
coloring of the neighborhood complex $\mathcal{N}(G)$ have been
proved by Civan and Yal\c{c}in \cite{Civan}.

\medskip
\noindent {\bf Definition 2.1.} Let $G$ be a graph and let $v \in
V(G)$. The {\it clique set} of a vertex $v$ is the set of all
maximal cliques of $G$ containing $v$ and is denoted by
$\mathcal{C}_G(v)$.

\medskip
\noindent {\bf Definition 2.2.} Let $G$ be a graph. A surjective
map $\mathcal{\kappa}: V(G) \rightarrow [k]$ is called a {\it
$k$-linear coloring} of $G$ if the collection $\{\mathcal{C}_G(v):
\mathcal{\kappa}(v)=i\}$ is linearly ordered by inclusion for all
$i \in [k]$, where $\mathcal{C}_G(v)$ is the clique set of $v$,
or, equivalently, for two vertices $v, u \in V(G)$, if
$\mathcal{\kappa}(v)=\mathcal{\kappa}(u)$ then either
$\mathcal{C}_G(v) \subseteq \mathcal{C}_G(u)$ or $\mathcal{C}_G(v)
\supseteq \mathcal{C}_G(u)$. The least integer $k$ for which $G$
is $k$-linear colorable is called the {\it linear chromatic
number} of $G$ and is denoted by $\lambda(G)$.

\subsection{Properties}

Next, we study the linear coloring on graphs and its association
to the proper vertex coloring. In particular, we show that for any
graph $G$ the linear chromatic number of $\skew3\overline{G}$ is
an upper bound for $\chi(G)$.

\medskip
\noindent {\bf Proposition 2.1.} {\it Let $G$ be a graph. If
$\mathcal{\kappa}: V(G) \rightarrow [k]$ is a $k$-linear coloring
of $\skew3\overline{G}$, then $\mathcal{\kappa}$ is a coloring of
the graph~$G$.}

\yy \noindent {\sl Proof.} \s Let $G$ be a graph and let
$\mathcal{\kappa}: V(G) \rightarrow [k]$ be a $k$-linear coloring
of $\skew3\overline{G}$. From Definition 2.2, we have that for any
two vertices $v, u \in V(G)$, if
$\mathcal{\kappa}(v)=\mathcal{\kappa}(u)$ then either
$\mathcal{C}_{\skew3\overline{G}}(v) \subseteq
\mathcal{C}_{\skew3\overline{G}}(u)$ or
$\mathcal{C}_{\skew3\overline{G}}(v) \supseteq
\mathcal{C}_{\skew3\overline{G}}(u)$ holds. Without loss of
generality, assume that $\mathcal{C}_{\skew3\overline{G}}(v)
\subseteq \mathcal{C}_{\skew3\overline{G}}(u)$ holds. Consider a
maximal clique $C \in \mathcal{C}_{\skew3\overline{G}}(v)$. Since,
$\mathcal{C}_{\skew3\overline{G}}(v) \subseteq
\mathcal{C}_{\skew3\overline{G}}(u)$, then $C \in
\mathcal{C}_{\skew3\overline{G}}(u)$. Thus, both $u, v \in C$ and
therefore $uv \in E(\skew3\overline{G})$ and $uv \notin E(G)$.
Hence, any two vertices assigned the same color in a $k$-linear
coloring of $\skew3\overline{G}$ are not neighbors in $G$.
Concluding, any $k$-linear coloring of $\skew3\overline{G}$ is a
coloring of $G$. \s \qed

\yy It is therefore straightforward to conclude the following.

\medskip
\noindent {\bf Corollary 2.1.} {\it For any graph $G$,
$\mathcal{\lambda}(\skew3\overline{G}) \geq \mathcal{\chi}(G)$.}

\begin{figure}[]
\y \hrule \y\y
  \centering
    \includegraphics[scale=0.7]{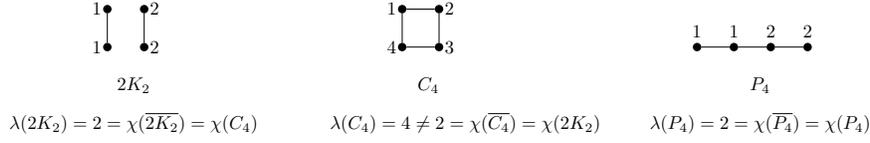}
  \centering
  \caption{Illustrating a linear coloring of the graphs $2K_2$, $C_4$ and
  $P_4$ with the least possible colors.} \label{graph3-4}
 \yy \hrule \y
\end{figure}

\yy In Figure~\ref{graph3-4} we depict a linear coloring of the
well known graphs $2K_2$, $C_4$ and $P_4$, using the least
possible colors, and show the relation between the chromatic
number $\mathcal{\chi}(G)$ of each graph $G \in \{2K_2, C_4,
P_4\}$ and the linear chromatic number
$\mathcal{\lambda}(\skew3\overline{G})$.

\medskip
\noindent {\bf Proposition 2.2.} {\it Let $G$ be a graph. A
coloring $\mathcal{\kappa}:V(G) \rightarrow [k]$ of $G$ is a
$k$-linear coloring of $\skew3\overline{G}$ if and only if either
$N_G(u) \subseteq N_G(v)$ or $N_G(u) \supseteq N_G(v)$ holds in
$G$, for every $u,v \in V(G)$ with
$\mathcal{\kappa}(u)=\mathcal{\kappa}(v)$.}

\yy \noindent {\sl Proof.} \s Let $G$ be a graph and let
$\mathcal{\kappa}: V(G) \rightarrow [k]$ be a coloring of $G$.
Assume that $\mathcal{\kappa}$ is a $k$-linear coloring of
$\skew3\overline{G}$. We will show that either $N_G(u) \subseteq
N_G(v)$ or $N_G(u) \supseteq N_G(v)$ holds in $G$ for every $u,v
\in V(G)$ with $\mathcal{\kappa}(u)=\mathcal{\kappa}(v)$. Consider
two vertices $v, u \in V(G)$, such that
$\mathcal{\kappa}(u)=\mathcal{\kappa}(v)$. Since
$\mathcal{\kappa}$ is a linear coloring of $\skew3\overline{G}$
then, from Definition 2.2, either
$\mathcal{C}_{\skew3\overline{G}}(u) \subseteq
\mathcal{C}_{\skew3\overline{G}}(v)$ or
$\mathcal{C}_{\skew3\overline{G}}(u) \supseteq
\mathcal{C}_{\skew3\overline{G}}(v)$ holds. Without loss of
generality, assume that $\mathcal{C}_{\skew3\overline{G}}(u)
\subseteq \mathcal{C}_{\skew3\overline{G}}(v)$. We will show that
$N_G(u) \supseteq N_G(v)$ holds in $G$. Assume the contrary. Thus,
a vertex $z \in V(G)$ exists, such that $z \in N_G(v)$ and $z
\notin N_G(u)$ and, thus, $zu \in E(\skew3\overline{G})$ and $zv
\notin E(\skew3\overline{G})$. Now consider a maximal clique $C$
in $\skew3\overline{G}$ which contains $z$ and $u$. Since $zv
\notin E(\skew3\overline{G})$ then $v \notin C$. Thus, there
exists a maximal clique $C$ in $\skew3\overline{G}$ such that $C
\in \mathcal{C}_{\skew3\overline{G}}(u)$ and $C \notin
\mathcal{C}_{\skew3\overline{G}}(v)$, which is a contrast to our
assumption that $\mathcal{C}_{\skew3\overline{G}}(u) \subseteq
\mathcal{C}_{\skew3\overline{G}}(v)$. Therefore, $N_G(u) \supseteq
N_G(v)$ holds in $G$.

\y Let $G$ be a graph and let $\mathcal{\kappa}: V(G) \rightarrow
[k]$ be a coloring of $G$. Assume now that either $N_G(u)
\subseteq N_G(v)$ or $N_G(u) \supseteq N_G(v)$ holds in $G$, for
every $u,v \in V(G)$ with
$\mathcal{\kappa}(u)=\mathcal{\kappa}(v)$. We will show that the
coloring $\mathcal{\kappa}$ of $G$ is a $k$-linear coloring of
$\skew3\overline{G}$. Without loss of generality, assume that
$N_G(u) \supseteq N_G(v)$ holds in $G$. We will show that
$\mathcal{C}_{\skew3\overline{G}}(u) \subseteq
\mathcal{C}_{\skew3\overline{G}}(v)$. Assume the opposite. Thus, a
maximal clique $C$ exists in $\skew3\overline{G}$, such that $C
\in \mathcal{C}_{\skew3\overline{G}}(u)$ and $C \notin
\mathcal{C}_{\skew3\overline{G}}(v)$. Now consider a vertex $z \in
V(G)$ ($z \neq u$ and $z \neq v$), such that $z \in C$ and $zv
\notin E(\skew3\overline{G})$. Such a vertex exists since $C$ is
maximal in $\skew3\overline{G}$ and $C \notin
\mathcal{C}_{\skew3\overline{G}}(v)$. Thus, $zv \notin
E(\skew3\overline{G})$ and $zu \in E(\skew3\overline{G})$. Hence,
$zv \in E(G)$ and $zu \notin E(G)$, which is a contrast to our
assumption that $N_G(u) \supseteq N_G(v)$. \s \qed

\yy Taking into consideration Definition 2.2 and Proposition 2.2,
we show the following.

\medskip
\noindent {\bf Corollary 2.2.} {\it Let $G$ be a graph and let
$\mathcal{\kappa}: V(G) \rightarrow [k]$ be a $k$-linear coloring
of $\skew3\overline{G}$. For every pair of vertices $u, v \in
V(G)$ for which $\mathcal{\kappa}(u)=\mathcal{\kappa}(v)$, the
following statements are equivalent:

\begin{itemize}

\item [(i)] $\mathcal{C}_{\skew3\overline{G}}(u) \subseteq
\mathcal{C}_{\skew3\overline{G}}(v)$ or
$\mathcal{C}_{\skew3\overline{G}}(u) \supseteq
\mathcal{C}_{\skew3\overline{G}}(v)$

\item [(ii)] $N_G(v) \subseteq N_G(u)$ or $N_G(v) \supseteq
N_G(u)$

\item [(iii)] $N_{\skew3\overline{G}}[u] \subseteq
N_{\skew3\overline{G}}[v]$ or $N_{\skew3\overline{G}}[u] \supseteq
N_{\skew3\overline{G}}[v]$.

\end{itemize}}

\noindent {\sl Proof.} \s From Definition 2.2 and Proposition 2.2,
it is easy to see that (i) $\Leftrightarrow$ (ii) holds. What is
left to show is (ii) $\Leftrightarrow$ (iii), which is
straightforward from basic set theory principles; specifically,
take into consideration that $N_G(u)=V(G) \backslash
N_{\skew3\overline{G}}[u]$, where $N_G(u)$ denotes the open
neighborhood of $u$ in $G$ and $N_{\skew3\overline{G}}[u]$ denotes
the closed neighborhood of $u$ in $\skew3\overline{G}$. \s \qed

\medskip
\noindent {\bf Observation 2.1.} It is easy to see that using
Corollary~2.2, the definition of a linear coloring of a graph $G$
can be restated as follows: A coloring $\mathcal{\kappa}: V(G)
\rightarrow [k]$ is a $k$-linear coloring of $G$ if the collection
$\{N_G[v]: \mathcal{\kappa}(v)=i\}$ is linearly ordered by
inclusion for all $i \in [k]$. Equivalently, for two vertices $v,
u \in V(G)$, if $\mathcal{\kappa}(v)=\mathcal{\kappa}(u)$ then
either $N_G[v] \subseteq N_G[u]$ or $N_G[v] \supseteq N_G[u]$.

\subsection{A Linear Coloring Algorithm}

In this section we present a polynomial time algorithm for linear
coloring which can be applied to any graph $G$, and provides an
upper bound for $\chi(G)$. Although we have introduced linear
coloring through Definition~2.2, in our algorithm we exploit the
property stated in Observation~2.1, since the problem of finding
all maximal cliques of a graph $G$ is not polynomially solvable on
general graphs. Before describing our algorithm, we first
construct a directed acyclic graph (DAG) $D_G$ of a graph $G$,
which we call {\it DAG associated to the graph $G$}, and we use it
in the proposed algorithm.


\yy \noindent {\bf The DAG $D_G$ associated to the graph $G$.} Let
$G$ be a graph. We first compute the closed neighborhood $N_G[v]$
of each vertex $v$ of $G$, and then, we construct the following
directed acyclic graph $D$, which depicts all inclusion relations
among the vertices' closed neighborhoods: $V(D)=V(G)$ and
$E(D)=\{\overrightarrow{xy} \s : \s x,y \in V(D) \s and \s N_G[x]
\subseteq N_G[y]\}$, where $\overrightarrow{xy}$ is a directed
edge from $x$ to $y$. In the case where the equality $N_G[x] =
N_G[y]$ holds, we choose to add one of the two edges so that the
resulting graph $D$ is acyclic (for example, we can use the
labelling of the vertices, and if $x<y$ then we add
$\overrightarrow{xy}$). It is easy to see that $D$ is a transitive
directed acyclic graph. Indeed, by definition $D$ is constructed
on a partially ordered set of elements $(V(D),\leq)$, such that
for some $x,y \in V(D)$, $x \leq y \Leftrightarrow N_G[x]
\subseteq N_G[y]$.

\y For reasons of simplicity, we consider the vertices of $D$
located in levels. In the first level we consider the vertices
with indegree equal to zero. For every vertex $y$ belonging to
level $\ell$ there exists at least one vertex $x$ in level
$\ell-1$ such that $\overrightarrow{xy}$. For every edge
$\overrightarrow{xy}$, if $x$ belongs to level $i$ and $y$ belongs
to level $j$, then $i<j$. For example, in the case where the
equality $N_G[x] = N_G[y]$ holds, and vertices $x$ and $y$ are
already located in levels $i$ and $j$ respectively, such that
$i<j$, then we choose to add the edge $\overrightarrow{xy}$.


\yy \noindent {\bf The algorithm for linear coloring.} Given a
graph $G$, the proposed algorithm computes a linear coloring and
the linear chromatic number of $G$. The algorithm works as
follows:

\begin{itemize}

\item [(i)] {\tt compute} the closed neighborhood set of every
vertex of $G$, and, then, find the inclusion relations among the
neighborhood sets and construct the DAG $D_G$ associated to the
graph $G$.

\item [(ii)] {\tt find} a minimum path cover $\mathcal{P}(D_G)$,
and its size $\mathcal{\rho}(D_G)$, of the transitive DAG $D_G$
(e.g. see \cite{Boesch77}).

\item [(iii)] {\tt assign} one color $\kappa(v)$ to each vertex $v \in
V(D_G)$, such that vertices belonging to the same path of
$\mathcal{P}(D_G)$ are assigned the same color and vertices of
different paths are assigned different colors; this is a
surjective map
$\mathcal{\kappa}:V(D_G)\rightarrow[\mathcal{\rho}(D_G)]$.

\item [(iv)] {\tt return} the value $\kappa(v)$ for each vertex $v \in
V(D_G)$ and the size $\mathcal{\rho}(D_G)$ of the minimum path
cover of $D_G$; $\kappa$ is a linear coloring of $G$ and
$\mathcal{\rho}(D_G)$ equals the linear chromatic number
$\mathcal{\lambda}(G)$ of $G$.

\end{itemize}

\noindent {\bf Correctness of the algorithm.} Let $G$ be a graph
and let $D_G$ be the DAG associated to the graph $G$. The
computation of a minimum path cover in a transitive DAG $D$ is
known to be polynomially solvable; the problem is equivalent to
the maximum matching problem in a bipartite graph formed from $D$
\cite{Boesch77}. Consider the value $\kappa(v)$ for each vertex $v
\in V(D_G)$ returned by the algorithm and the size
$\mathcal{\rho}(D_G)$ of a minimum path cover of $D_G$. We show
that the surjective map
$\mathcal{\kappa}:V(D_G)\rightarrow[\mathcal{\rho}(D_G)]$ is a
linear coloring of the vertices of $G$, and prove that the size
$\mathcal{\rho}(D_G)$ of the minimum path cover $\mathcal{P}(D_G)$
of the DAG $D_G$ is equal to the linear chromatic number
$\mathcal{\lambda}(G)$ of the graph $G$.

\medskip
\noindent {\bf Proposition 2.3.} {\it Let $G$ be a graph and let
$D_G$ be the DAG associated to the graph $G$. A path cover of
$D_G$ gives a linear coloring of the graph $G$ by assigning a
particular color to all vertices of each path. Moreover, the size
$\mathcal{\rho}(D_G)$ of the minimum path cover $\mathcal{P}(D_G)$
of the graph~$D_G$ equals to the linear chromatic number
$\lambda(G)$ of the graph~$G$.}

\yy \noindent {\sl Proof.} \s Let $G$ be a graph, $D_G$ be the DAG
associated to $G$, and let $\mathcal{P}(D_G)$ be a minimum path
cover of $D_G$. The size $\mathcal{\rho}(D_G)$ of the DAG $D_G$,
equals to the minimum number of directed paths in $D_G$ needed to
cover the vertices of $D_G$ and, thus, the vertices of $G$. Now,
consider a coloring $\kappa:V(D_G)\rightarrow[k]$ of the vertices
of $D_G$, such that vertices belonging to the same path are
assigned the same color and vertices of different paths are
assigned different colors. Therefore, we have
$\mathcal{\rho}(D_G)$ colors and $\mathcal{\rho}(D_G)$ sets of
vertices, one for each color. For every set of vertices belonging
to the same path, their corresponding closed neighborhood sets can
be linearly ordered by inclusion. Indeed, consider a path in $D_G$
with vertices $\{v_1, v_2,\ldots,v_m\}$ and edges
$\overrightarrow{v_iv_{i+1}}$ for $i \in \{1,2,\ldots,m\}$. From
the construction of $D_G$, it holds that $\forall i,j \in
\{1,2,\ldots,m\}$, $\overrightarrow{v_iv_j} \in E(D_G)$
$\Leftrightarrow$ $N_G[v_i] \subseteq N_G[v_j]$. In other words,
the corresponding neighborhood sets of the vertices belonging to a
path in $D_G$ are linearly ordered by inclusion. Thus, the
coloring $\kappa$ of the vertices of $D_G$ gives a linear coloring
of $G$. This linear coloring $\kappa$ is optimal, uses
$k=\mathcal{\rho}(D_G)$ colors, and gives the linear chromatic
number $\mathcal{\lambda}(G)$ of the graph $G$. Indeed, suppose
that there exists a different linear coloring
$\kappa':V(D_G)\rightarrow[k']$ of $G$ using $k'$ colors, such
that $k' < k$. For every color given in $\kappa'$, consider a set
consisted of the vertices assigned that color. It is true that for
the vertices belonging to the same set, their neighborhood sets
are linearly ordered by inclusion. Therefore, these vertices can
belong to the same path in $D_G$. Thus, each set of vertices in
$G$ corresponds to a path in $D_G$ and, additionally, all vertices
of $G$ (and therefore of $D_G$) are covered. This is a path cover
of $D_G$ of size $\mathcal{\rho}'(D_G)=k' <
k=\mathcal{\rho}(D_G)$, which is a contradiction since
$\mathcal{P}(D_G)$ is a minimum path cover of $D_G$. Therefore, we
conclude that the linear coloring
$\kappa:V(D_G)\rightarrow[\mathcal{\rho}(D_G)]$ is optimal, and
hence, $\mathcal{\rho}(D_G)=\mathcal{\lambda}(G)$. \s \qed

\section{Co-linear Graphs}

In Section~2 we showed that for any graph $G$, the linear
chromatic number $\lambda(\skew3\overline{G})$ of
$\skew3\overline{G}$ is an upper bound for the chromatic number
$\chi(G)$ of $G$, i.e. $\chi(G) \leq \lambda(\skew3\overline{G})$.
Recall that a known lower bound for the chromatic number of $G$ is
the clique number $\omega(G)$ of $G$, i.e. $\chi(G) \geq
\omega(G)$. Motivated by the Perfect Graph Theorem \cite{Gol}, in
this section we exploit our results on linear coloring and we
study those graphs for which the equality $\chi(G) =
\lambda(\skew3\overline{G})$ holds for every induce subgraph. The
outcome of this study was the definition of a new class of perfect
graphs, namely co-linear graphs. We also prove structural
properties for its members.

\medskip
\noindent {\bf Definition 3.1.} A graph $G$ is called {\it
co-linear} if and only if
$\mathcal{\chi}(G_A)=\mathcal{\lambda}(\skew3\overline{G}_A)$,
$\forall A \subseteq V(G)$; a graph $G$ is called {\it linear} if
$\skew3\overline{G}$ is a co-linear graph.

\yy Next, we show that co-linear graphs are perfect; actually, we
show that they form a subclass of the class of co-chordal graphs,
a superclass of the class of threshold graphs and they are
distinguished from the class of split graphs. We first give some
definitions and show some interesting results.

\medskip
\noindent {\bf Definition 3.2.} The edge $uv$ of a graph $G$ is
called {\it actual} if neither $N_G[u] \subseteq N_G[v]$ nor
$N_G[u] \supseteq N_G[v]$. The set of all actual edges of $G$ will
be denoted by $E_{\alpha}(G)$.

\medskip
\noindent {\bf Definition 3.3.} A graph $G$ is called {\it
quasi-threshold} if it has no induced subgraph isomorphic to a
$C_4$ or a $P_4$ or, equivalently, if it contains no actual edges.

\yy More details on actual edges and characterizations of
quasi-threshold graphs through a classification of their edges can
be found in \cite{Nik00}. The following result directly follows
from Definition 3.2 and Corollary 2.2.

\medskip
\noindent {\bf Proposition 3.1.} {\it Let $\mathcal{\kappa}: V(G)
\rightarrow [k]$ be a $k$-linear coloring of the graph $G$. If the
edge $uv \in E(G)$ is an actual edge of $G$, then
$\mathcal{\kappa}(u) \neq \mathcal{\kappa}(v)$.}

\yy Based on Definitions~3.1 and 3.2, and Proposition 3.1, we
prove the following result.

\medskip
\noindent {\bf Proposition 3.2.} {\it Let $G$ be a graph and let
$F$ be the graph such that $V(F)=V(G)$ and $E(F)=E(G) \cup
E_{\alpha}(\skew3\overline{G})$. The graph $G$ is a co-linear
graph if and only if $\mathcal{\chi}(G_A)=\mathcal{\omega}(F_A)$,
$\forall A \subseteq V(G)$.}

\yy \noindent {\sl Proof.} \s Let $G$ be a graph and let $F$ be a
graph such that $V(F)=V(G)$ and $E(F)=E(G) \cup
E_{\alpha}(\skew3\overline{G})$, where
$E_{\alpha}(\skew3\overline{G})$ is the set of all actual edges of
$\skew3\overline{G}$. From Definition 3.1, $G$ is a co-linear
graph if and only if $\mathcal{\chi}(G_A) =
\mathcal{\lambda}(\skew3\overline{G}_A)$, $\forall A \subseteq
V(G)$. It suffices to show that
$\mathcal{\lambda}(\skew3\overline{G}_A) = \mathcal{\omega}(F_A)$,
$\forall A \subseteq V(G)$. From Corollary 2.2, it is easy to see
that two vertices which are not connected by an edge in
$\skew3\overline{G}_A$ belong necessarily to different cliques,
and thus, they cannot receive the same color in a linear coloring
of $\skew3\overline{G}_A$. In other words, the vertices which are
connected by an edge in $G_A$ cannot take the same color in a
linear coloring of $\skew3\overline{G}_A$. Moreover, from
Proposition 3.1 vertices which are endpoints of actual edges in
$\skew3\overline{G}_A$ cannot take the same color in a linear
coloring of $\skew3\overline{G}_A$.

\y Next, we construct the graph $F_A$ with vertex set
$V(F_A)=V(G_A)$ and edge set $E(F_A)=E(G_A) \cup
E_{\alpha}(\skew3\overline{G}_A)$, where
$E_{\alpha}(\skew3\overline{G}_A)$ is the set of all actual edges
of $\skew3\overline{G}_A$. Every two vertices in $F_A$, which have
to take a different color in a linear coloring of
$\skew3\overline{G}_A$ are connected by an edge. Thus, the size of
the maximum clique in $F_A$ equals to the size of the maximum set
of vertices which pairwise must take a different color in
$\skew3\overline{G}_A$, i.e. $\mathcal{\omega}(F_A) =
\mathcal{\lambda}(\skew3\overline{G}_A)$ holds for all $A
\subseteq V(G)$. Concluding, $G$ is a co-linear graph if and only
if $\mathcal{\chi}(G_A)=\mathcal{\omega}(F_A)$, $\forall A
\subseteq V(G)$. \s \qed

\yy Taking into consideration Proposition 3.2 and the structure of
the edge set $E(F)=E(G) \cup E_{\alpha}(\skew3\overline{G})$ of
the graph $F$, it is easy to see that $E(F)=E(G)$ if
$\skew3\overline{G}$ has no actual edges. Actually, this will be
true for all induced subgraphs, since if $G$ is a quasi-threshold
graph then $G_A$ is also a quasi-threshold graph for all $A
\subseteq V(G)$. Thus,
$\mathcal{\chi}(G_A)=\mathcal{\omega}(F_A)$, $\forall A \subseteq
V(G)$. Therefore, the following result holds.

\medskip
\noindent {\bf Corollary 3.1.} {\it Let $G$ be a graph. If
$\skew3\overline{G}$ is quasi-threshold, then $G$ is a co-linear
graph.}

\yy From Corollary 3.1 we obtain a more interesting result.

\medskip
\noindent {\bf Proposition 3.3} {\it Any threshold graph is a
co-linear graph.}

\yy \noindent {\sl Proof.} \s Let $G$ be a threshold graph. It has
been proved that an undirected graph $G$ is a threshold graph if
and only if $G$ and its complement $\skew3\overline{G}$ are
quasi-threshold graphs \cite{Nik00}. From Corollary 3.1, if
$\skew3\overline{G}$ is quasi-threshold then $G$ is a co-linear
graph. Concluding, if $G$ is threshold, then $\skew3\overline{G}$
is quasi-threshold and thus $G$ is a co-linear graph. \s \qed

\yy However, not any co-linear graph is a threshold graph. Indeed,
Chv\'{a}tal and Hammer \cite{Chv77} showed that threshold graphs
are $(2K_2,P_4,C_4)$-free, and, thus, the graphs $P_4$ and $C_4$
are co-linear graphs but not threshold graphs (see
Figure~\ref{graph3-4}). We note that the proof that any threshold
graph $G$ is a co-linear graph can be also obtained by showing
that any coloring of a threshold graph $G$ is a linear coloring of
$\skew3\overline{G}$ by using Proposition 2.2, Corollary 2.1 and
the property that $N(u) \subseteq N[v]$ or $N(v) \subseteq N[u]$
for any two vertices $u, v$ of $G$. However, Proposition 3.2 and
Corollary 3.1 actually give us a stronger result since the class
of quasi-threshold graphs is a superclass of the class of
threshold graphs.

\y The following result is even more interesting, since it places
the class of co-linear graphs into the map of perfect graphs as a
subclass of co-chordal graphs.

\medskip
\noindent {\bf Proposition 3.4.} {\it Any co-linear graph is a
co-chordal graph.}

\yy \noindent {\sl Proof.} \s Let $G$ be a co-linear graph. It has
been showed that a co-chordal graph is $(2K_2, antihole)$-free
\cite{Gol}. To show that any co-linear graph $G$ is a co-chordal
graph we will show that if $G$ has a $2K_2$ or an $antihole$ as
induced subgraph, then $G$ is not a co-linear graph. Since by
definition a graph $G$ is co-linear if and only if the equality
$\mathcal{\chi}(G_A) = \mathcal{\lambda}(\skew3\overline{G}_A)$
holds for every induced subgraph $G_A$ of $G$, it suffices to show
that the graphs $2K_2$ and $antihole$ are not co-linear graphs.

\y The graph $2K_2$ is not a co-linear graph, since
$\mathcal{\chi}(2K_2)=2 \neq 4 = \mathcal{\lambda}(C_4)$; see
Figure~\ref{graph3-4}. Now, consider the graph
$G=\skew3\overline{C}_n$ which is an antihole of size $n \geq 5$.
We will show that $\mathcal{\chi}(G) \neq
\mathcal{\lambda}(\skew3\overline{G})$. It follows that
$\mathcal{\lambda}(\skew3\overline{G})=\mathcal{\lambda}(C_n)=n
\geq 5$, i.e. if the graph $\skew3\overline{G}=C_n$ is to be
colored linearly, every vertex has to take a different color.
Indeed, assume that a linear coloring $\mathcal{\kappa}: V(G)
\rightarrow [k]$ of $\skew3\overline{G}=C_n$ exists such that for
some $u_i, u_j \in V(G)$, $i \neq j$, $1 \leq i, j \leq n$,
$\mathcal{\kappa}(u_i)=\mathcal{\kappa}(u_j)$. Since $u_i, u_j$
are vertices of a hole, their neighborhoods in
$\skew3\overline{G}$ are $N[u_i]=\{u_{i-1},u_i, u_{i+1}\}$ and
$N[u_j]=\{u_{j-1},u_j, u_{j+1}\}$, $2 \leq i, j \leq n-1$. For
$i=1$ or $i=n$, $N[u_1]= \{u_{n}, u_{2}\}$ and $N[u_n]= \{u_{n-1},
u_{1}\}$. Since $\mathcal{\kappa}(u_i)=\mathcal{\kappa}(u_j)$,
from Corollary 2.2 we obtain that one of the inclusion relations
$N[u_i] \subseteq N[u_j]$ or $N[u_i] \supseteq N[u_j]$ must hold
in $\skew3\overline{G}$. Obviously this is possible if and only if
$i=j$, for $n \geq 5$; this is a contradiction to the assumption
that $i \neq j$. Thus, no two vertices in a hole take the same
color in a linear coloring. Therefore,
$\mathcal{\lambda}(\skew3\overline{G})=n$. It suffices to show
that $\mathcal{\chi}(G) < n$. It is easy to see that for the
antihole $\skew3\overline{C}_n$, $deg(u)=n-3$, for every vertex $u
\in V(G)$. Brook's theorem \cite{Brooks} states that for an
arbitrary graph $G$ and for all $u \in V(G)$, $\mathcal{\chi}(G)
\leq max \{d(u)+1\} = (n-3)+1=n-2$. Therefore, $\mathcal{\chi}(G)
\leq n-2 < n = \mathcal{\lambda}(\skew3\overline{G})$. Thus the
antihole $\skew3\overline{C}_n$ is not a co-linear graph.

\y We have showed that the graphs $2K_2$ and $antihole$ are not
co-linear graphs. It follows that any co-linear graph is $(2K_2,
antihole)$-free and, thus, any co-linear graph is a co-chordal
graph. \s \qed

\begin{figure}[]
\y \hrule \y\y
  \centering
    \includegraphics[scale=0.6]{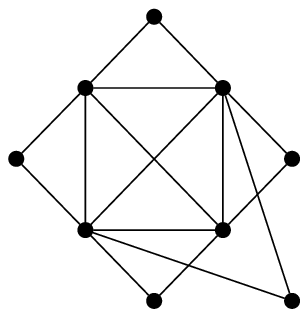}
  \centering
  \caption{A graph $G$ which is a split graph but not co-linear,
  since $\mathcal{\chi}(G)=4$ and $\mathcal{\lambda}(\skew3\overline{G})=5$.} \label{graph1}

\yy

  \centering
    \includegraphics[scale=0.7]{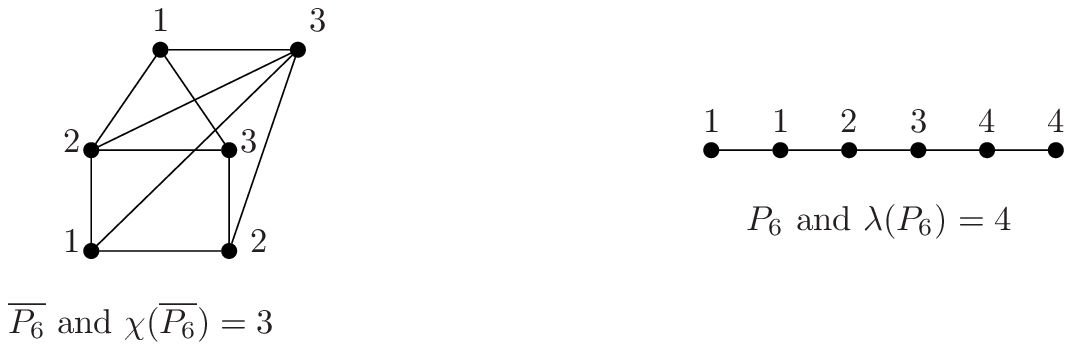}
  \centering
  \caption{Illustrating the graph $\skew3\overline{P}_6$ which is not a co-linear graph,
  since $\mathcal{\chi}(\skew3\overline{P}_6) \neq \mathcal{\lambda}(P_6)$.} \label{graph2}
 \yy \hrule \y
\end{figure}

\yy Although any co-linear graph is co-chordal, the reverse is not
always true. For example, the graph $G$ in Figure~\ref{graph1} is
a co-chordal graph but not a co-linear graph. Indeed,
$\mathcal{\chi}(G)=4$ and
$\mathcal{\lambda}(\skew3\overline{G})=5$. It is easy to see that
this graph is also a split graph. Moreover, the class of split
graphs is distinguished from the class of co-linear graphs since
the graph $C_4$ is a co-linear graph but not a split graph, and
the graph $G$ in Figure~\ref{graph1} is a split graph but not a
co-linear graph. However, the two classes are not disjoint; an
example is the graph $C_3$. Recall that a graph $G$ is a {\it
split graph} if there is a partition of the vertex set $V(G)=K+I$,
where $K$ induces a clique in $G$ and $I$ induces an independent
set; split graphs are characterized as $(2K_2,C_4,C_5)$-free
graphs.


\y We have proved that co-linear graphs are $(2K_2,
antihole)$-free. Note that, since $\overline{C_5}=C_5$ and also
the chordless cycle $C_n$ is $2K_2$-free for $n \ge 6$, it is easy
to see that co-linear graphs are $hole$-free. In addition,
$\skew3\overline{P}_6$ is another forbidden induced subgraph for
co-linear graphs (see Figure~\ref{graph2}). Thus, we obtain the
following result.

\medskip
\noindent {\bf Proposition 3.5.} {\it If $G$ is a co-linear graph,
then $G$ is $(2K_2, antihole, \skew3\overline{P}_6)$-free.}

\yy The forbidden graphs $2K_2$, $antihole$, and
$\skew3\overline{P}_6$ are not enough to characterize completely
the class of co-linear graphs, since split graphs do not contain
any of these graphs as an induced subgraph. Thus, split graphs
which are not co-linear graphs cannot be characterized by these
forbidden induced subgraphs; see Figure~\ref{graph1}.

\section{Linear Graphs}

In this section we study the complement class of co-linear graphs,
namely linear graphs, in terms of forbidden induced subgraphs, and
we derive inclusion relations between the class of linear graphs
and other classes of perfect graphs.

\subsection{Properties}

\y We first provide a characterization of linear graphs by means
of linear coloring on graphs. Since co-linear graphs are perfect,
it follows that if $G$ is a co-linear graph
$\chi(G_A)=\omega(G_A)=\alpha(\skew3\overline{G_A})$, $\forall A
\subseteq V(G)$. Therefore, the following characterization of
linear graphs holds.

\medskip
\noindent {\bf Proposition 4.1.} {\it A graph $G$ is linear if and
only if $\mathcal{\alpha}(G_A)=\mathcal{\lambda}(G_A)$, $\forall A
\subseteq V(G)$.}

\yy From Corollary~2.1 and Proposition~4.1 we obtain the following
characterization for linear graphs.

\medskip
\noindent {\bf Proposition 4.2.} {\it Linear graphs are those
graphs $G$ for which the linear chromatic number achieves its
theoretical lower bound in every induced subgraph of $G$.}

\yy Directly from Corollary~3.1 we can obtain the following
result: any quasi-threshold graph is a linear graph. From
Propositions~3.5 and 4.1 we obtain that linear graphs are $(C_4,
hole, P_6)$-free. Therefore, the following result holds.

\medskip
\noindent {\bf Proposition 4.3.} {\it Any linear graph is a
chordal graph.}

\yy Although any linear graph is chordal, the reverse is not
always true, i.e. not any chordal graph is a linear graph. For
example, the complement $\skew3\overline{G}$ of the graph
illustrated in Figure~\ref{graph1} is a chordal graph but not a
linear graph. Indeed, $\alpha(\skew3\overline{G})=4$ and
$\lambda(\skew3\overline{G})=5$. It is easy to see that this graph
is also a split graph. Moreover, the class of split graphs is
distinguished from the class of linear graphs since the graph
$2K_2$ is a linear graph but not a split graph, and the graph
$\skew3\overline{G}$ of Figure~\ref{graph1} is a split graph but
not a linear graph. However, the two classes are not disjoint; an
example is the graph $C_3$.

\y Another known subclass of the class of chordal graphs is the
class of strongly chordal graphs. The following definitions and
results given by Farber \cite{Far83} turn up to be useful in
proving some results about the structure of linear graphs. More
details about strongly chordal graphs can be found in
\cite{Brand99, Far83}.

\begin{figure}[]
\y \hrule \yy
  \centering
    \includegraphics[scale=0.6]{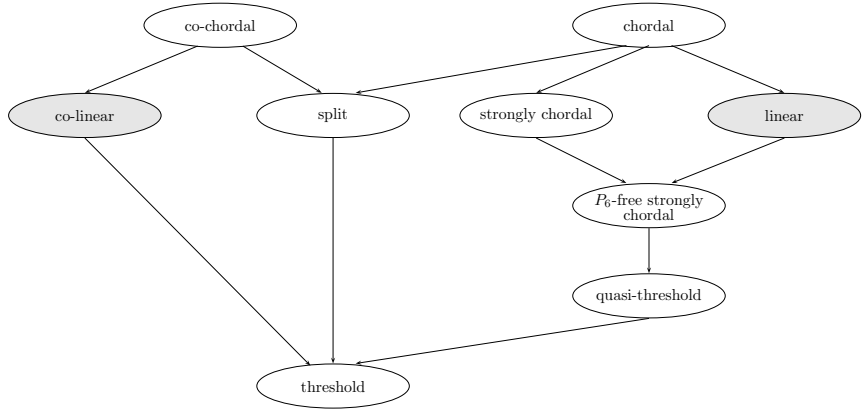}
  \centering
  \caption{Illustrating the inclusion relations among the classes
  of linear graphs, co-linear graphs, and other classes of
  perfect graphs.} \label{classes-both}
 \yy \hrule \y
\end{figure}

\medskip
\noindent {\bf Definition 4.2.} (Farber \cite{Far83}) A vertex
ordering $(v_1, v_2, \dots, v_n)$ is a {\it strong perfect
elimination ordering} of a graph $G$ iff $\sigma$ is a perfect
elimination ordering and also has the property that for each $i$,
$j$, $k$ and $\ell$, if $i<j$, $k<\ell$, $v_k,v_\ell \in N[v_i]$,
and $v_k \in N[v_j]$, then $v_\ell \in N[v_j]$. A graph is {\it
strongly chordal} iff it admits a strong perfect elimination
ordering.

\medskip
\noindent {\bf Definition 4.3.} (Farber \cite{Far83}) Let $G$ be a
graph. A vertex $v$ is {\it simple} in $G$ if $\{N[x]: x \in
N[v]\}$ is linearly ordered by inclusion.

\medskip
\noindent {\bf Theorem 4.1.} (Farber \cite{Far83}) {\it A graph
$G$ is strongly chordal if and only if every induced subgraph of
$G$ has a simple vertex.}

\medskip
\noindent {\bf Corollary 4.1.} (Chang \cite{Chang88}) {\it A
strong perfect elimination ordering of a graph $G$ is a vertex
ordering $(v_1, v_2, \dots, v_n)$ such that for all $i \in
\{1,2,\ldots,n\}$ the vertex $v_i$ is simple in $G_i$ and also
$N_{G_i}[v_\ell] \subseteq N_{G_i}[v_k]$ whenever $i \leq \ell
\leq k$ and $v_\ell, v_k \in N_{G_i}[v_i]$.}

\yy The following characterization of strongly chordal graphs will
be next used to derive properties about the structure of linear
graphs. We first give the following definition.

\medskip
\noindent {\bf Definition 4.1.} An {\it incomplete $k$-sun} $S_k$
($k \geq 3$) is a chordal graph on $2k$ vertices whose vertex set
can be partitioned into two sets, $U=\{u_1,u_2,\ldots,u_k\}$ and
$W=\{w_1,w_2,\ldots,w_k\}$, so that $W$ is an independent set, and
$w_i$ is adjacent to $u_j$ if and only if $i=j$ or $i=j+1$ (mod
$k$). A {\it $k$-sun} is an incomplete $k$-sun $S_k$ in which $U$
is a complete graph.

\medskip
\noindent {\bf Proposition 4.4.} (Farber \cite{Far83}) {\it A
chordal graph $G$ is strongly chordal if and only if it contains
no induced $k$-sun.}

\subsection{Forbidden Subgraphs}

\y Hereafter, we study the structure of the forbidden induced
subgraphs of the class of linear graphs, and we prove that any
$P_6$-free chordal graph which is not a linear graph properly
contains a $k$-sun as an induced subgraph.

\y We consider the class of $P_6$-free chordal graphs which we
have shown that it properly contains the class of linear graphs.
Let $\mathcal{F}$ be the family of all the minimal forbidden
induced subgraphs of the class of linear graphs. Let $F_i$ be a
member of $\mathcal{F}$, which is neither a $C_n$ ($n \geq 4$) nor
a $P_6$. We next prove the main result of this section: any graph
$F_i$ properly contains a $k$-sun ($k \geq 3$) as an induced
subgraph. From Proposition~4.4 it suffices to show that any
$P_6$-free strongly chordal graph is a linear graph and also that
the $k$-sun ($k \geq 3$) is a linear graph.

\y Let $G$ be a $P_6$-free strongly chordal graph. In order to
show that $G$ is a linear graph we will show that
$\alpha(G)=\lambda(G)$ and that the equality holds for every
induced subgraph of $G$. Let $L$ be the set of all simple vertices
of $G$, and $S$ be the set of all simplicial vertices of $G$; note
that $L \subseteq S$ since a simple vertex is also a simplicial
vertex. First, we construct a maximum independent set $I$ and a
strong perfect elimination ordering $\sigma$ of $G$ with special
properties needed for our proof. Next, we assign a coloring
$\kappa:V(G)\rightarrow [k]$ to the vertices of $G$, where
$k=\alpha(G)=|I|$, and show that $\kappa$ is an optimal linear
coloring of $G$. Actually, we show that we can assign a linear
coloring with $\lambda(G)=\alpha(G)$ colors to any $P_6$-free
strongly chordal graph, by using the constructed strong perfect
elimination ordering $\sigma$ of $G$. Finally, we show that the
equality $\lambda(G_A)=\alpha(G_A)$ holds for every induced
subgraph $G_A$ of $G$.

\yy \noindent {\bf Construction of $I$ and $\sigma$.} Let $G$ be a
$P_6$-free strongly chordal graph, and let $L$ be the set of all
simple vertices in $G$. From Definition~4.2, $G$ admits a strong
perfect elimination ordering. Using a modified version of the
algorithm given by Farber in \cite{Far83} we construct a strong
perfect elimination ordering $\sigma=(v_1,v_2,\ldots,v_n)$ of the
graph $G$ having specific properties. Our algorithm also
constructs the maximum independent set $I$ of $G$. Since $G$ is a
chordal graph and $\sigma$ is a perfect elimination ordering, we
can use a known algorithm (e.g. see \cite{Gol}) to compute a
maximum independent set of the graph $G$. Throughout the
algorithm, we denote by $G_i$ the subgraph of $G$ induced by the
set of vertices $V(G) \backslash \{v_1,v_2,\ldots,v_{i-1}\}$,
where $v_1,v_2,\ldots,v_{i-1}$ are the vertices which have already
been added to the ordering $\sigma$ during the construction.
Moreover, we denote by $I^*$ the set of vertices which have not
been added to $\sigma$ yet and additionally do not have a neighbor
already added in $\sigma$ which belongs to $I$.

\y In Figure \ref{algorithm}, we present a modified version of the
algorithm given by Farber \cite{Far83} for constructing a strong
perfect elimination ordering $\sigma$ of $G$. Our algorithm in
each iteration of Steps 3--5 adds to the ordering $\sigma$ all
vertices which are simple in $G_i$, while Farber's algorithm
selects only one simple vertex of $G_i$ and adds it to $\sigma$.
We note that $L_i$ is the set of all the simple vertices of $G_i$
and $v_i$ is that vertex of $L_i$ which is added first to the
ordering $\sigma$. It is easy to see that the constructed ordering
$\sigma$ is a strong perfect elimination ordering of $G$, since
every vertex which is simple in $G$ is also simple in every
induced subgraph of $G$. Clearly, the constructed set $I$ is a
maximum independent set of $G$.

\begin{figure}[t]


\yy \hrule \yyy \noindent{\it Input:} a strongly chordal graph
$G$; \y \noindent{\it Output:} a strong perfect elimination
ordering $\sigma$ of $G$;

\begin{itemize}

\item[1.] {\tt set} $I=\emptyset$, $I^*=V(G)$, $\sigma=\emptyset$,
$n=|V(G)|$, and $V_0=V(G)$;

\item[2.] Let $(V_0,<_0)$ be the partial ordering on $V_0$ in which $v <_0 u$ if and only if $v=u$.\\
{\tt set} $V_1=V(G)$ and $i=1$;

\item[3.] Let $G_i$ be the subgraph of $G$ induced by $V_i$, that is, $V_i=V(G_i)$. \\
{\tt construct} an ordering on $V_i$ by $v <_i u$ if $v <_{i-1} u$ or $N_i[v] \subset N_i[u]$;\\
{\tt set} $k=i$;

\item[4.] Let $L_k$ be the set of all the simple vertices
in $G_i$.

{\tt while} $L_k \neq \emptyset$ {\tt do}\\
        \mytab $\circ$ {\tt construct} an ordering on $V_i$ by $v <_i u$ if $v <_{i-1} u$ or $N_i[v] \subset N_i[u]$; \\
        \mytab \phantom{$\circ$} {\tt choose} a vertex $v_i$ which belongs to $L_k$ and is minimal in $(V_i,<_i)$ to add to the ordering;\\
        \mytab \phantom{$\circ$} {\tt set} $V_{i+1}=V_i \backslash \{v_i\}$ and $L_k=L_k \backslash
        \{v_i\}$;\\
        \mytab $\circ$ {\tt if} $v_i \in I^*$ {\tt then}\\
        \mytab \mytab \phantom{$\circ$} {\tt set} $I=I \cup \{v_i\}$ and $I^* = I^* \backslash
                \{v_i\}$;\\
        \mytab \mytab \phantom{$\circ$} {\tt delete} all neighbors of $v_i$ from $I^*$;\\
        \mytab $\circ$ {\tt set} $i=i+1$;\\
{\tt end-while;}

\item[5.] {\tt if} \,$i=n+1$ \, {\tt then} output the ordering $\sigma = (v_1, v_2, \ldots, v_n)$ of $V(G)$ and {\tt stop};\\
\mytab \mytab \mytab \mytab {\tt else} go to step 3;

\end{itemize}

\caption{A modified version of Farber's algorithm for constructing
a strong perfect elimination ordering $\sigma$ and a maximum
independent set $I$ of a strongly chordal graph~$G$.}
\label{algorithm}

\yyy \hrule

\end{figure}

\y From the fact that $G$ is a $P_6$-free strongly chordal graph
and from the construction of $I$ and $\sigma$ we obtain the
following properties.

\yy \noindent {\bf Property 4.1.} Let $G$ be a $P_6$-free strongly
chordal graph and let $L$ be the set of all simple vertices of
$G$. For each vertex $v_x \notin L$, there exists a chordless path
of length at most 4 connecting $v_x$ to any vertex $v \in L$.

\medskip\noindent {\bf Property 4.2.} Let $G$ be a $P_6$-free strongly
chordal graph, $L$ be the set of all simple vertices of $G$, and
let $I$ and $\sigma$ be the maximum independent set and the
ordering, respectively, constructed by our algorithm. Then,
\begin{itemize}

\item[(i)] if $v_i \notin L$ and $i<j$, then $v_{j} \notin L$;

\item[(ii)] for each vertex $v_x \notin I$, there exists a
vertex $v_i \in I$, $i<x$, such that $v_x \in N_{G_i}[v_i]$.

\end{itemize}

\y Next, we describe an algorithm for assigning a coloring
$\kappa$ to the vertices of $G$ using exactly $\alpha(G)$ colors
and, then, we show that $\kappa$ is a linear coloring of $G$.

\yy \noindent {\bf The coloring $\kappa$ of $G$.} Let $G$ be a
$P_6$-free strongly chordal graph, and let $L$ (resp. $S$) be the
set of all simple (resp. simplicial) vertices in $G$. We consider
a maximum independent set $I$, and a strong elimination ordering
$\sigma$, as constructed above. Now, in order to compute the
linear chromatic number $\lambda(G)$ of $G$, we assign a coloring
$\kappa$ to the vertices of $G$ and show that $\kappa$ is a linear
coloring of $G$. Actually, we show that we can assign a linear
coloring with $\lambda(G)=\alpha(G)$ colors to any $P_6$-free
strongly chordal graph, by using the constructed strong perfect
elimination ordering $\sigma$ of $G$.

\y\noindent First, we assign a coloring $\kappa:V(G)\rightarrow
[k]$, where $k=\alpha(G)$, to the vertices of $G$ as follows:

\begin{itemize}

\item[1. ] Successively visit the vertices in the ordering
$\sigma$ from left to right, and color the first vertex $v_i \in
I$ which has not been assigned a color yet, with color
$\kappa(v_i)$.

\item[2. ] Color all uncolored vertices $v_k \in N_{G_i}(v_i)$, with
color $\kappa(v_k)=\kappa(v_i)$.

\item[3. ] Repeat steps 1 and 2 until there are no
uncolored vertices $v_i \in I$ in $G$.

\end{itemize}

\y Based on this process, we obtain that every vertex $v_i$
belonging to the maximum independent set $I$ of $G$ is assigned a
different color in step~1, and for each such vertex $v_i$ all its
uncolored neighbors to its right in the ordering $\sigma$ are
assigned the same color with $v_i$ in step~2. Therefore, so far we
have assigned $\alpha(G)$ colors to the vertices of $G$. Now, from
Property 4.2(ii) it is easy to see that $\kappa$ is a coloring of
the vertex set $V(G)$, i.e. there is no vertex in $\sigma$ which
has not been assigned a color. Thus, $\kappa$ is a coloring of $G$
using $\alpha(G)$ colors. Note that $\kappa$ is not a proper
vertex coloring of $G$. Actually, since the following lemma holds,
from Proposition 2.1 it appears that $\kappa$ is a proper vertex
coloring of $\skew3\overline{G}$.

\medskip
\noindent {\bf Lemma 4.1.} {\it The coloring $\kappa$ is a linear
coloring of $G$.}

\yy \noindent {\sl Proof.} \s Let $G$ be a $P_6$-free strongly
chordal graph, and let $L$ (resp. $S$) be the set of all simple
(resp. simplicial) vertices in $G$. We consider a maximum
independent set $I$, a strong elimination ordering $\sigma$, and a
coloring $\kappa$ of $G$, as constructed above. Hereafter, for two
vertices $v_i$ and $v_j$ in the ordering $\sigma$, we say that
$v_i<v_j$ if the vertex $v_i$ appears before the vertex $v_j$ in
$\sigma$.

\y Next, we show that $\kappa$ is a linear coloring of $G$, that
is, the collection $\{\mathcal{C}_G(v_\ell): \kappa(v_\ell)=j\}$
is linearly ordered by inclusion for all $j \in [k]$. From
Corollary~2.2, it is equivalent to show that the collection
$\{N_G[v_\ell]: \kappa(v_\ell)=j\}$ is linearly ordered by
inclusion for all $j \in [k]$. Each such collection contains
exactly one set $N_G[v_i]$ where $v_i \in I$, and some sets
$N_G[v_k]$ where $v_k$ are neighbors of $v_i$ in $G_i$ and
$\kappa(v_k)=\kappa(v_i)$. Thus, it suffices to show that for each
vertex $v_i \in I$, the collection $\{N_G[v_k]: v_k \in
N_{G_i}[v_i] \s and \s \kappa(v_k)=\kappa(v_i) \}$ is linearly
ordered by inclusion. To this end, we distinguish two cases
regarding the vertices $v_i \in I$; in the first case we consider
$v_i$ to be a simplicial vertex, that is $v_i \in S$, and in the
second case we consider $v_i \notin S$.

\medskip \noindent {\bf Case 1: The vertex $v_i \in I$ and $v_i \in S$.} Since
$\sigma$ is a strong elimination ordering, each vertex $v_i \in I$
is simple in $G_i$ and thus $\{N_{G_i}[v_k]: v_k \in
N_{G_i}[v_i]\}$ is linearly ordered by inclusion. We will show
that $\{N_G[v_k]: v_k \in N_{G_i}[v_i] \s and \s
\kappa(v_k)=\kappa(v_i) \}$ is linearly ordered by inclusion for
all vertices $v_i \in I \cap S$. Recall that in the coloring
$\kappa$ of $G$ we assign the color $\kappa(v_k)=\kappa(v_i)$ to a
vertex $v_k \notin I$, if $v_i \in I$, $v_k \in N_{G_i}[v_i]$ and
there exists no vertex $v_{i'} \in I$ such that $v_k \in
N_{G_{i'}}[v_{i'}]$ and $v_i'<v_i$ in $\sigma$. By definition, if
$v_i \in L$ then the collection $\{N_G[v_k]: v_k \in N_{G_i}[v_i]
\s and \s \kappa(v_k)=\kappa(v_i) \}$ is linearly ordered by
inclusion. Thus, hereafter we consider vertices $v_i \in I \cap S$
and $v_i \notin L$.

\y Consider that the vertex $v_i$ has a neighbor $v_1$ to its left
in the ordering $\sigma$, i.e. $v_1 < v_i$. Since $v_i$ is a
simplicial vertex in $G$, its closed neighborhood forms a clique
and, thus, $v_1 \in N_G[v_k]$ for all vertices $v_k \in
N_{G_i}[v_i]$. Therefore, the existence of such a vertex $v_1$
preserves the linear order by inclusion of $\{N_{G_i}[v_k] \cup
\{v_1\}: v_k \in N_{G_i}[v_i]\}$. Thus, $N_G[v_i] \subseteq
N_G[v_k]$, for all vertices $v_k \in N_{G_i}[v_i]$ and
$\kappa(v_k)=\kappa(v_i)$.

\y Now, consider that the vertex $v_i$ has two neighbors $v_k$ and
$v_j$ to its right in the ordering $\sigma$, such that
$v_i<v_k<v_j$ and $\kappa(v_k)=\kappa(v_j)=\kappa(v_i)$; thus,
$N_{G_i}[v_k] \subseteq N_{G_i}[v_j]$. In the case where the
equality $N_{G_i}[v_k] = N_{G_i}[v_j]$ holds, without loss of
generality, we may assume that the degree of $v_k$ in $G$ is less
than or equal to the degree of $v_j$ in $G$ (note that $\sigma$ is
still a strong elimination ordering). Assume that $N_G[v_k]
\subseteq N_G[v_j]$ does not hold. Then, there exist vertices
$v_2$ and $v_3$ in $G$ such that $v_2 \in N_G[v_k]$, $v_2 \notin
N_G[v_j]$, $v_3 \in N_G[v_j]$, and $v_3 \notin N_G[v_k]$. Since
$N_{G_i}[v_k] \subseteq N_{G_i}[v_j]$, it is easy to see that
$v_2<v_i$ in $\sigma$. Assume that $v_2$ is the first (from left
to right) neighbor of $v_k$ in $\sigma$. Since
$\kappa(v_k)=\kappa(v_i)$, it follows that $v_2 \notin I$.
Moreover, from Property 4.2(ii) it holds that there exists a
vertex $v_4 \in I$, such that $v_4 < v_2$ and $v_2 \in N_G[v_4]$.
Additionally, since $\kappa(v_k)=\kappa(v_j)=\kappa(v_i)$ it holds
that $v_k, v_j \notin N_G[v_4]$. Hence, the subgraph of $G$
induced by the vertices $\{v_4,v_2,v_k,v_j\}$ is a $P_4$.
Concerning now the position of the vertex $v_3$ in the ordering
$\sigma$, we can have either $v_3 < v_i$ in the case where
$N_{G_i}[v_k] = N_{G_i}[v_j]$ holds, or $v_3
> v_i$ otherwise. We will show that in both cases we are leaded to a
contradiction to our initial assumptions; that is, either it
results that $G$ has a $P_6$ as an induced subgraph or that the
vertices should be added to $\sigma$ in an order different to the
one originally assumed.

\medskip\noindent {\bf Case 1.1.} $v_3 < v_i$. It is easy to see
that $v_3 \notin I$, since otherwise $v_j$ would have taken the
color $\kappa(v_j)=\kappa(v_3)$ during the coloring $\kappa$ of
$G$. Thus, from Property 4.2(ii) there exists a vertex $v_5 \in
I$, such that $v_5 < v_3$ and $v_3 \in N_G[v_5]$. Therefore, the
vertices $\{v_4, v_2, v_k, v_j, v_3, v_5\}$ induce a $P_6$ in $G$,
which is also chordless since $G$ is chordal.

\medskip\noindent {\bf Case 1.2.} $v_3 > v_i$.
Since $v_i \notin L$, from Property 4.2(i) it follows that $v_3
\notin L$. Thus, from Property 4.1 we obtain that there exists a
chordless path of length at most 4 connecting $v_3 \notin L$ to
any vertex $v \in L$. Similarly, it easily follows that $v_4 \in
L$. However, we know that in a non-trivial strongly chordal graph
there exist at least two non adjacent simple vertices
\cite{Far83}. Thus, there exist a vertex $v \in L$, $v \neq v_4$,
such that the distance $d(v,v_3)$ of $v_3$ from $v$ is at most 4.
Let $d_m(v_3,v)=max\{d(v_3,v) : \s \forall v \in L, \s v \neq v_4
\}$. Since $v_3 \notin L$ and $G$ is $P_6$-free, it follows that
$1 \leq d_m(v,v_3) \leq 4$.

\y Next, we distinguish four cases regarding the maximum distance
$d_m(v_3,v)$ and show that each one comes to a contradiction. In
each case we have that $\{v_4, v_2, v_k, v_j, v_3\}$ is a
chordless path on five vertices. We first explain what is
illustrated in Figures~\ref{cases1} and ~\ref{cases2}. Let $G_y$
be the induced subgraph of $G$, such that during the construction
of $\sigma$ the vertex $v_i$ is simple in $G_y$, i.e. $v_i \in
L_y$ and $v_y \leq v_i$. In the two figures, the vertices are
placed on the horizontal dotted line in the order that appear in
the ordering $\sigma$. For the vertices which are not placed on
the dotted line, we are only interested about illustrating the
edges among them. The vertices which are to the right of the
vertical dashed line belong to the induced subgraph $G_y$ of $G$.
The dashed edges illustrate edges that may or may not exist in the
specific case. Next, we distinguish the four cases, and show that
each one of them comes to a contradiction:

\begin{figure}[t]
\hrule \medskip \hspace*{0.6in}

    \begin{minipage}[l]{2in}
    \includegraphics[scale=0.8]{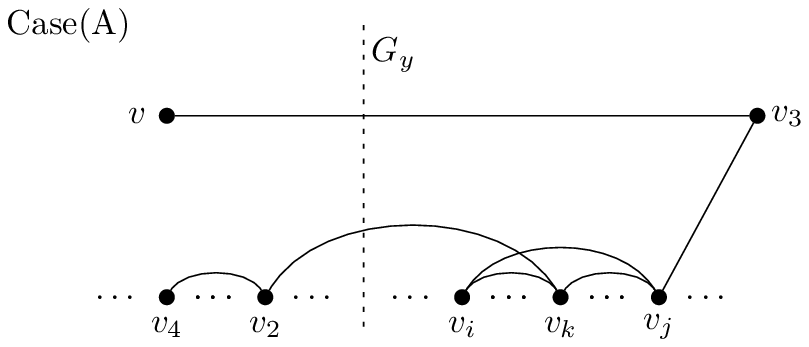}
    \end{minipage}

\vspace{0.2in} \hspace*{0.6in}

    \begin{minipage}[r]{2in}
    \includegraphics[scale=0.8]{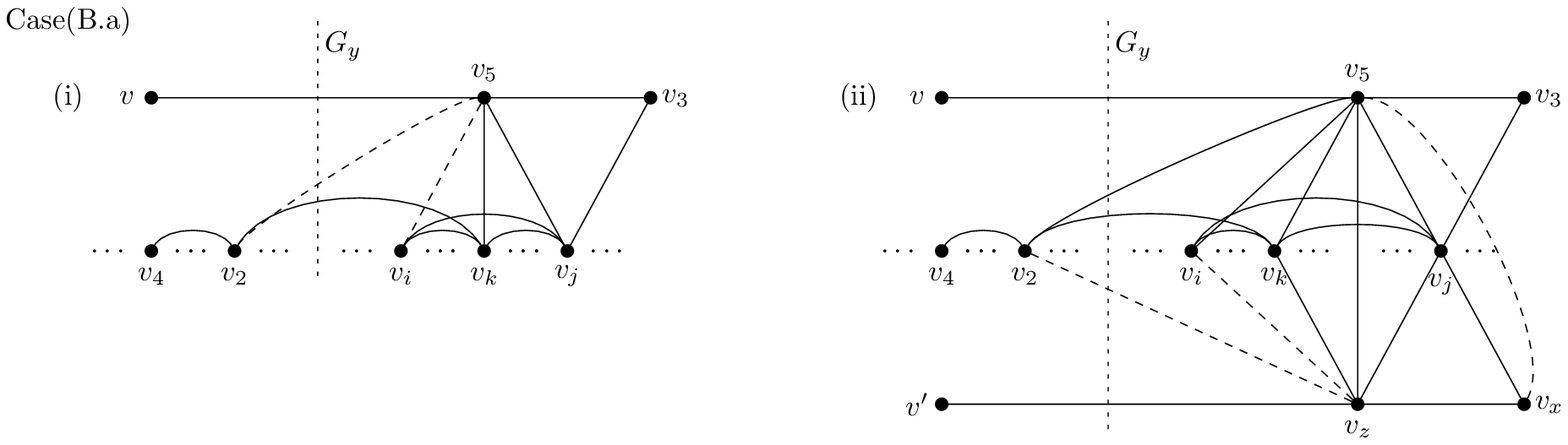}
    \end{minipage}

\vspace{0.2in}
    \caption{\small{Illustrating Case (A) and Case (B.a)}}
\medskip \hrule
 \label{cases1}
\end{figure}

\begin{itemize}

    \item[] {\bf Case (A):} $d_m(v_3,v)=1$.
    \y It is easy to see that $v_jv \notin E(G)$,
    since otherwise $v_j$ would have been assigned the color
    $\kappa(v)$ and not $\kappa(v_i)$ as assumed. Thus, in this case
    there exists a $P_6$ in $G$ induced by the vertices $\{v_4,
    v_2, v_k, v_j, v_3, v\}$; since $G$ is
    a chordal graph, other
    edges among the vertices of this path do not exist. This is a contradiction to our assumption that
    $G$ is a $P_6$-free graph.

\newpage

    \item[] {\bf Case (B):} $d_m(v_3,v)=2$.
    \y In this case there exists a vertex $v_5$ such that $\{v_3,
    v_5,v\}$ is a chordless path from $v_3$ to $v$.
    It follows that there exists a $P_7$ induced by
    the vertices $\{v_4, v_2, v_k, v_j, v_3, v_5,
    v\}$. Having assumed that $G$ is a $P_6$-free graph, the path $\{v_4, v_2, v_k, v_j, v_3\}$ is chordless
    and $v_j, v_k \notin N_G[v]$, we obtain that $v_jv_5 \in E(G)$ and $v_kv_5 \in
    E(G)$. Next, we distinguish three cases regarding the neighborhood of
    the vertex $v_3$ in $G$ and show that each one comes to a contradiction.

    \begin{itemize}

    \item[(B.a)] The vertex $v_3$ does not have neighbors in
    $G$ other than $v_5$ and $v_j$. In Case (i) we examine the cases
    where either $v_2v_5 \notin E(G)$ or $v_2v_5 \in E(G)$ and $v_j$ does not
    have a neighbor $v_x$ in $G_i$, such that $v_xv_k \notin
    E(G)$. In Case (ii) we examine the case where $v_2v_5 \in E(G)$ and $v_j$ has a
    neighbor $v_x$ in $G_i$, such that $v_xv_k \notin E(G)$.

        \begin{itemize}

        \item[(i)] Assume that $v_2v_5 \notin E(G)$. In this case, we
        can see that during the construction of $\sigma$, after the
        first iteration where $v$ and $v_4$ are added in the ordering, the
        vertex $v_3$ becomes simple in the remaining induced
        subgraph of $G$, since $N[v_5]$ becomes a subset of $N[v_j]$. Thus, $v_3$ can be added to $\sigma$ during the second
        iteration of the algorithm, along with $v_2$. However,
        $v_i$ will not be added to the ordering before the third
        iteration, since $v_i$ is not simple before $v_2$ is added to
        $\sigma$. Thus, we conclude that $v_3$ will be added in $\sigma$ before
        $v_i$, and more specifically that $v_3 < v_y \leq v_i$, and this is a contradiction to our assumption that $v_3 > v_i$.

        \y Now, assume that $v_2v_5 \in E(G)$. We know that $v_2$ is simple in the subgraph $G_2$ of $G$ induced
        by the vertices to the right of $v_2$ in $\sigma$. If $v_5, v_k \in N_{G_2}[v_2]$, $v_3 \in N_{G_2}[v_5]$, and
        $v_3 \notin N_{G_2}[v_k]$, then $N_{G_2}[v_5] \supset N_{G_2}[v_k]$. More specifically,
        since we have assumed that $v_2$ is the
        first (from left to right) neighbor of $v_k$ in $\sigma$, it follows that $N_G[v_5] \supset
        N_G[v_k]$. We know that $N_{G_i}[v_k] \subset N_{G_i}[v_j]$, and since we have assumed that $v_j$ does not have
        a neighbor $v_x$, such that $v_x < v_i$, it easily follows
        that $N_{G_i}[v_k] \subset N_{G_i}[v_j]=N_G[v_j]$. Thus, for every neighbor
        of $v_j$ in $G$, which is also a neighbor of $v_k$, we obtain that it is a
        neighbor of $v_5$ as well.

        \y Therefore, in the case where $v_j$ does not
        have a neighbor $v_x$ in $G$, and thus in $G_i$, such that $v_xv_k \notin
        E(G)$, it follows that $N_G[v_5]$ is a superset of $N_G[v_j]$ and, thus, the
        vertex $v_3$ is simple in $G$. Again we conclude that $v_3$
        will be added to $\sigma$ before $v_i$, and more specifically that $v_3 < v_y \leq v_i$. This is a
        contradiction to our assumption that $v_3 > v_i$.

        \item[(ii)] Consider now the case where $v_2v_5 \in E(G)$ and $v_j$
        has a neighbor $v_x$ in $G$, and thus in $G_i$,
        such that $v_xv_k \notin E(G)$. We will show that in this case either $v_3$ is simple
        after the first iteration, i.e. $v_x \in N[v_5]$ or $v_x$
        becomes simple after the first iteration. Since $v_x > v_i$ it follows
        that $v_x \notin L$. Therefore, there exists a path in $G$ from $v_x$
        to a vertex $v' \in L$ of length $d(v_x,v')$ at most
        4. Consider the case where $d(v_x,v')=1$. If $v \equiv v'$, then $v_5v_x \in E(G)$, since $G$ is a
        chordal graph; thus, $N[v_5] \supseteq N[v_j]$ and $v_3 \in L$. It is easy to see that $v' \neq v_4$, since
        $G$ is a chordal graph. Therefore, in the case where $v'v_x \in E(G)$, the graph $G$
        has a $P_6$ induced by the vertices $\{v_4, v_2, v_k, v_j, v_x,
        v'\}$. Thus, $v'v_x \notin E(G)$ and there exists a vertex
        $v_z$ such that $\{v_x, v_z, v'\}$ is a chordless path from
        $v_x$ to $v'$. Therefore, there exists a $P_7$ in $G$ and, thus, $v_k, v_j \in
        N_G[v_z]$. Additionally, from Case(B.a)(i) we have that $v_5 \in N_G[v_z]$
        (recall that if $v_2v_5 \in E(G)$, then $N_G[v_5] \supset N_G[v_k]$).

        \y Note that, the vertices $v_x$ and $v_z$ play the same role in $G$ as the vertices $v_3$ and
        $v_5$, respectively. Therefore, in the case where $v_2v_z \notin E(G)$, the vertex $v_x$ is
        simple after the first iteration and will be added to $\sigma$
        during the second iteration, while $v_i$ will be added during the third. Thus, we will have $v_x < v_y <
        v_i$ which is a contradiction to our assumption that $v_x >
        v_i$. Consider now the case where $v_2v_z \in E(G)$. Since $v_2$ is simple
        in the subgraph $G_2$ of $G$ induced
        by the vertices to the right of $v_2$ in $\sigma$, we must have either $v_zv_3 \in E(G)$ or $v_5v_x \in E(G)$.
        Without loss of generality assume that $v_5v_x \in E(G)$.
        Concluding, we have shown that even in the case where $v_j$ has a neighbor $v_x$ in $G$, and thus in $G_i$,
        such that $v_xv_k \notin E(G)$, then $N_G[v_5]$ is a
        superset of $N_G[v_j]$, and thus $v_3 \in L$. Thus, we have again $v_3 < v_y <
        v_i$ which is a contradiction to our assumption that $v_3 >
        v_i$. The same holds even if, additionally to the other edges, $v_4v_5 \in E(G)$.

        \end{itemize}

    \end{itemize}

    \y So far, we have shown that if $v_3$ has the
    vertices $v_j$ and $v_5$ as neighbors, then either $v_3 \in L$ or $v_3$ is simple in the second
    iteration, that is before $v_i$ can be added to $\sigma$ (i.e.
    $v_3 < v_y \leq v_i$). This is due to the fact that for any
    neighbor $v_5$ of $v_3$ we have shown that $N[v_5] \subseteq
    N[v_j]$ in the case where $v_2v_5 \notin E(G)$, and $N[v_5] \supseteq
    N[v_j]$ in the case where $v_2v_5 \in E(G)$; thus $v_3$ will be added to $\sigma$ before $v_i$.
    Since we initially assumed that $v_3>v_i$ in $\sigma$, i.e. that $v_3$ does not become
    simple before $v_i$ becomes simple, we continue by examining the cases where $v_3$ has neighbors in
    $G_y$ other than $v_5$ and $v_j$.

    \begin{itemize}

    \item[(B.b)] The vertex $v_3$ has two neighbors $v_5$ and
        $v_5'$ in $G_y$, such that $v_5v_5' \notin E(G)$. Since we have assumed that the maximum distance of
        the vertex $v_3$ from $v$ in $G$, for any vertex $v \in L$, $v \neq v_4$, is
        $d_m(v_3,v)=2$, and $v_3$ has no neighbor belonging to $L$, it follows that $v_5, v_5' \notin L$ and
        there exist vertices $v, v' \in L$ such that the vertices $\{v_3, v_5 ,v\}$ induce a chordless path from $v_3$ to
        $v$ and $\{v_3, v_5' ,v'\}$ induce a chordless path from $v_3$ to $v'$.
        It is easy to see that $v \neq v'$ and $vv' \notin E(G)$ since $G$ is a chordal graph.
        Therefore, from Case (B.a) we have $v_k, v_j \in N_G[v_5]$
        and $v_k, v_j \in N_G[v_5']$. However, in this case there
        exists a $C_4$ in $G$ induced by the vertices $\{v_5, v_3, v_5',
        v_k\}$, since by assumption $v_5v_5' \notin E(G)$ and $v_3v_k \notin
        E(G)$. It easily follows that the same arguments hold for any two neighbors of $v_3$ in $G$.
        Concluding, the vertex $v_3$ cannot have two neighbors $v_5$ and
        $v_5'$ in $G$, such that $v_5v_5' \notin E(G)$. Thus, $v_3 \in
        S$.

\begin{figure}[]
\yy \hrule \y\y\y
  \centering
    \includegraphics[scale=0.8]{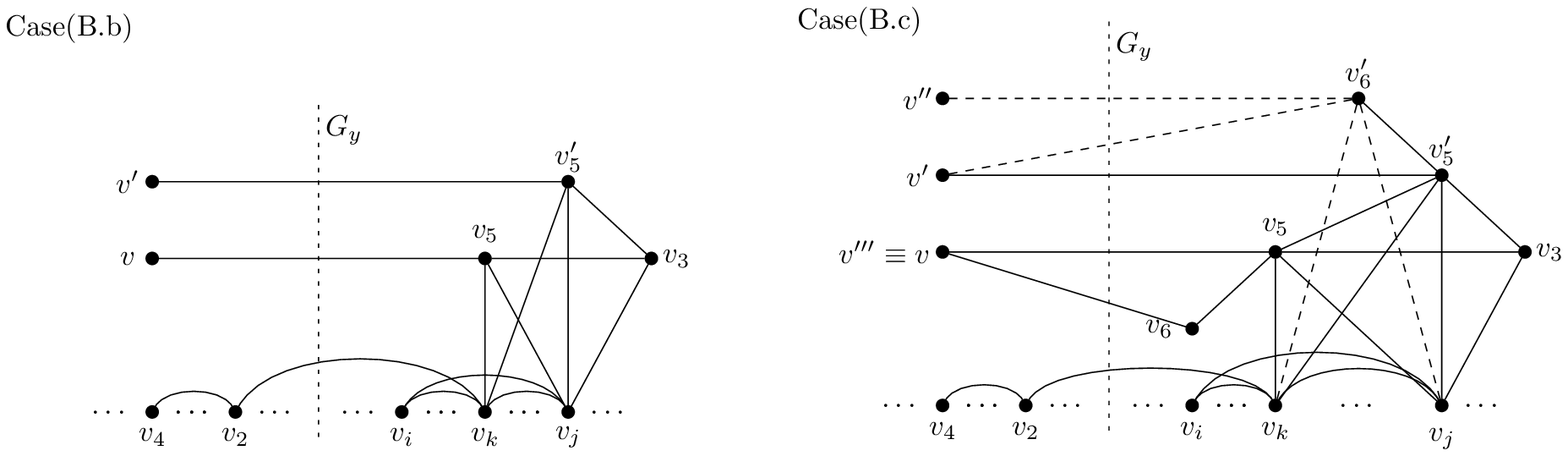}
  \centering
  \caption{Illustrating Cases (B.b) and (B.c) of the proof.} \label{cases2}
 \yy \hrule \y\y
\end{figure}

    \item[(B.c)] The vertex $v_3$ has two neighbors $v_5$ and
        $v_5'$ (where $v_5 \neq v_j$ and $v_5' \neq v_j$) in $G_y$,
        such that $v_5v_5' \in E(G)$, but neither $N_y[v_5] \subseteq N_y[v_5']$
        nor $N_y[v_5'] \subseteq N_y[v_5]$; thus, there exist
        vertices $v_6$ and $v_6'$ in $G_y$ such that $v_5v_6 \in E(G)$ and
        $v_5v_6' \notin E(G)$ and, also, $v_5'v_6' \in E(G)$ and
        $v_5'v_6 \notin E(G)$. Since $v_3 \in S$, it follows
        that $v_6, v_6' \notin N_G[v_3]$. Since $d_m(v_3,v)=2$,
        there exists a vertex $v \in L$ such that $\{v_3, v_5, v\}$ is a chordless path
        from $v_3$ to $v$. Similarly, there exists a vertex $v' \in L$ such that $\{v_3, v_5, v'\}$ is a chordless path
        from $v_3$ to $v'$. We have that $v \neq v'$, $vv_5' \notin E(G)$ and $v'v_5 \notin
        E(G)$, since otherwise $v$ and $v'$ would not be simple
        in $G$. Additionally, $vv' \notin E(G)$, $vv_6' \notin E(G)$, and $v'v_6 \notin E(G)$,
        since $G$ is a chordal graph. Therefore, from Case (B.a) we have $v_k, v_j \in N_G[v_5]$
        and $v_k, v_j \in N_G[v_5']$. Assume that there exist vertices $v'', v''' \in L$,
        such that $v_6v''' \in E(G)$ and $v_6'v'' \in E(G)$. It is
        easy to see that at least one of the equivalences $v \equiv v'''$ and
        $v' \equiv v''$ holds, otherwise $G$ has a $P_6$ induced by the vertices
        $\{v''', v_6, v_5, v_5', v_6', v''\}$. Without loss of generality, assume
        that $v \equiv v'''$ holds.

        \y Since $v \in L$, $v_5, v_6 \in N_G[v]$, $v_5' \in N_G[v_5]$, and $v_5' \notin
        N_G[v_6]$, it follows that $N_G[v_6] \subset N_G[v_5]$. In
        the case where $v_k, v_j \notin N_G[v_6]$ we have $v_6 \in L$ and, thus, $v_6$ would be
        added to $\sigma$ in the first iteration which is a
        contradiction to our assumption that $v_6 \in G_y$.
        Assume that $v_jv_6 \in E(G)$; it follows that $v_kv_6 \in E(G)$, since otherwise $G$ has a $P_6$
        induced by the vertices $\{v_4, v_2, v_k, v_j, v_6, v\}$. If $v' \equiv v''$, the same arguments
        hold for $v_6'$ too and, thus, if $v_jv_6' \in E(G)$ then $v_kv_6' \in E(G)$. In the case where $v' \neq v''$
        we have $v_6'v_k \in E(G)$, since
        otherwise $G$ has a $P_6$ induced by the vertices $\{v_4, v_2, v_k, v_5', v_6', v''\}$.
        Thus, in any case $v_6, v_6' \in N_G[v_k]$, and $G$ has a $3$-sun induced by the vertices
        $\{v_k, v_5, v_5', v_6', v_6, v_3\}$. Since other edges
        between the vertices of the $3$-sun do not exist, it
        follows that at least one of the vertices $v_6$ and $v_6'$
        does not belong to the neighborhood of $v_k$ and, thus, of
        $v_j$ in $G$. Without loss of generality, let $v_6$ be that vertex. Thus, $v_6 \in
        L$ and, subsequently, $v_6$ will be added to $\sigma$
        during the first iteration. Thus, $v_3$ is simple and will be added to
        $\sigma$ during the second iteration, along with $v_2$,
        while $v_i$ will be added to $\sigma$ after the second
        iteration (i.e. $v_3 < v_y \leq v_i$). This is a
        contradiction to our assumption that $v_3 > v_i$.

        \y Using similar arguments, we can prove that $v_3$ will be
        added to $\sigma$ before $v_i$, even if there exist edges between $v_2$ and the vertices $v_5$, $v_5'$, $v_6$, and
        $v_6'$. Actually, it easily follows that $v_2v_6 \notin
        E(G)$, since $v_6v_k \notin E(G)$ and $G$ is a chordal graph.
        Additionally, $v_2v_5 \notin E(G)$, since we know that $v_5v_6' \notin E(G)$, $v_kv_3 \notin E(G)$ and $v_2$
        is simple in $G_2$. Therefore, whether $v_2v_5', v_2v_6' \in
        E(G)$ or not, it does not change the fact that $v_3$ becomes simple after the first
        iteration and, thus, $v_3$ is added to $\sigma$ before $v_i$.
        Note, that even in the case where $v \equiv v_4$ or $v' \equiv v_4$,
        it similarly follows that $v_6' \in L$ or $v_6 \in L$ respectively
        and, thus, $v_3$ becomes simple after the first
        iteration and is added to $\sigma$ before $v_i$.

    \end{itemize}

    \item[] {\bf Case (C):} $d_m(v_3,v)=3$.
    \y In this case there exist vertices $v_5$ and $v_6$ such that $\{v_3,
    v_5, v_6, v\}$ is a chordless path from $v_3$ to $v$. Since
    now $G$ has a $P_8$, it follows that $v_5v_j \in E(G)$ and, additionally,
    some other edges must exist among the vertices $v_2$, $v_k$, $v_j$, $v_5$, and
    $v_6$. In any case, we will prove that either $N_G[v_5] \subseteq
    N_G[v_j]$ or $N_G[v_j] \subseteq N_G[v_5]$ and, thus, $v_3 \in L$.
    Similarly to Case (B), we distinguish three cases regarding the neighborhood of
    the vertex $v_3$ in $G$ and show that if $v_3 \notin L$ then each one comes to a contradiction.

        \begin{itemize}

        \item[(C.a)] The vertex $v_3$ does not have neighbors in $G$ other than $v_5$ and
        $v_j$. Consider the case where $v_3 \notin L$ because $v_6 \notin N_G[v_j]$ and $v_k \notin
        N_G[v_5]$. In this case, $G$ has a $P_7$ induced by the vertices $\{v_4, v_2, v_k, v_j, v_5, v_6, v\}$
        which is chordless since $G$ is a chordal
        graph; this is a contradiction to our assumption that $G$ is $P_6$-free. Consider, now, the case where $v_3 \notin L$
        because $v_6 \notin N_G[v_j]$ and $v_i \notin
        N_G[v_5]$. Since $G$ is $P_6$-free it follows that $v_5v_k \in
        E(G)$ and $v_6v_k \in E(G)$. However, in this case $G$ has
        a $3$-sun, unless either $v_iv_6 \in E(G)$ and, thus, $v_jv_6 \in
        E(G)$, or $v_iv_5 \in E(G)$. In either case it follows that $v_3 \in
        L$.

        \y Consider, now, the case where $v_j$ has another neighbor $v_x$ in $G_i$ such that $v_xv_5 \notin E(G)$.
        Using similar arguments as in Case (B.a)(ii), we come to a contradiction to our assumptions.
        More specifically, in the case where $v_2v_5 \in E(G)$, it is proved that $N_G[v_5] \supset
        N_G[v_j]$, and thus $v_3 \in L$. Similarly, in the case where $v_6v_j \notin E(G)$, it is proved
        that the vertex $v_x$ will be simple after the first iteration during the construction of $\sigma$, and thus $v_x < v_y \leq
        v_i$.

        \item[(C.b)] The vertex $v_3$ has two neighbors $v_5$ and $v_5'$ in $G_y$, such that $v_5v_5' \notin
        E(G)$. Using the same arguments as in Case (B.b), we obtain that in this case $G$ has a $C_4$ which
        is a contradiction to our assumptions.

        \item[(C.c)] The vertex $v_3$ has two neighbors $v_5$ and
        $v_5'$ (where $v_5 \neq v_j$ and $v_5' \neq v_j$) in $G_y$,
        such that $v_5v_5' \in E(G)$, and neither $N_y[v_5] \subseteq N_y[v_5']$
        nor $N_y[v_5'] \subseteq N_y[v_5]$; that is, there exist
        vertices $v_6$ and $v_6'$ in $G_y$ such that $v_5v_6 \in E(G)$ and
        $v_5v_6' \notin E(G)$ and, also, $v_5'v_6' \in E(G)$ and
        $v_5'v_6 \notin E(G)$. Similarly to Case (B.c), we can prove that this case comes to a contradiction as well. Note that, in this
        case $d_m(v_3,v)=3$ and, thus, there exists
        a chordless path $\{v_3, v_5, v_7, v\}$ from $v_3$ to $v$.
        Again, at least one of $v \equiv v'''$ and $v' \equiv v''$ must hold, since otherwise $G$
        has a $P_6$ induced by the vertices $\{v''', v_6, v_5, v_5', v_6',
        v''\}$. Using the same arguments as in Case (B.c), we obtain that if
        $v \equiv v'''$ then $v_k, v_j \notin N_G[v_6]$. However, now, we must additionally
        have $v_6v_7 \in E(G)$, since otherwise $G$ has a $C_4$
        induced by the vertices $\{v, v_7, v_5, v_6\}$. Therefore, as
        in Case (B.c) we obtain $v_6 \in L$, which is a
        contradiction to our assumption that the vertex $v_i$ appears in the ordering before the vertices $v_6$, $v_6'$, $v_5$, and $v_5'$.

        \end{itemize}

    \item[] {\bf Case (D):} $d_m(v_3,v)=4$.
    \y In this case there exist vertices $v_5$, $v_6$ and $v_7$ such that $\{v_3,
    v_5, v_6, v_7, v\}$ is a chordless path from $v_3$ to $v$. Since
    now $G$ has a $P_9$, it follows that $v_5v_j \in E(G)$ and, additionally,
    some other edges must exist. Similarly to Cases (A) and (B), we distinguish three cases regarding the neighborhood of
    the vertex $v_3$ in $G$ and show that if $v_3 \notin L$ then each one comes to a contradiction.

        \begin{itemize}

        \item[(D.a)] The $v_3$ does not have neighbors in $G$ other than $v_5$ and
        $v_j$. If we assume that $v_3 \notin L$, then $v_5$ has a
        neighbor in $G$ which is not a neighbor of $v_j$ and,
        additionally, $v_j$ has a neighbor in $G$ which is not a neighbor of
        $v_5$. Thus, we can have one of the following three cases, each of which comes to a contradiction:

            \begin{itemize}

            \item[$\bullet$] $v_2 \in N_G[v_5]$ and $v_7 \in
            N_G[v_j]$. Now, we have that $v_2v_6 \in E(G)$, since otherwise $G$ has a $P_6$
            induced by the vertices $\{v_4, v_2, v_5, v_6, v_7,
            v\}$. However, in this case $v_2$
            would not be simple in $G_2$, where $G_2$ is the subgraph of $G$ induced by
            the vertices to the right of $v_2$ in $\sigma$, since $v_7 \in N_G[v_6]$ and $v_7 \notin N_G[v_5]$
            and, also, $v_3 \in N_G[v_5]$ and $v_3 \notin N_G[v_6]$. Indeed, it
            suffices to show that the vertices $v_5$, $v_6$, $v_7$,
            and $v_3$ belong to the induced subgraph $G_2$ of $G$.

            \y We know that $v_5, v_3 \in N_G[v_j]$ and, thus, $v_5> v_i$ and $v_3 > v_i$ since
            we have assumed that $v_j$ does not have a neighbor
            $v_x$, such that $v_x < v_i$. Additionally, from $v_7 \in N_G[v_j]$ it follows that
            $v_6 \in N_G[v_j]$, since otherwise $G$ has a $C_4$ induced by the
            vertices $\{v_j, v_5, v_6, v_7\}$. Therefore, $v_6, v_7 \in
            N_G[v_j]$ and, thus, $v_i < v_6$ and $v_i < v_7$. Therefore, the vertices $v_5$, $v_6$, $v_7$,
            and $v_3$ belong to the induced subgraph $G_2$ of $G$,
            and thus, the vertex $v_2$ is not simple in $G_2$,
            which is a contradiction to our assumption that $\sigma$ is a strong perfect elimination ordering.

            \item[$\bullet$] $v_k \notin N_G[v_5]$ and $v_6 \notin
            N_G[v_j]$. From $v_k \notin N_G[v_5]$ we obtain that $v_2, v_i \notin N_G[v_5]$. In this case $G$ has a $P_8$ induced by the
            vertices $\{v_4, v_2, v_k, v_j, v_5, v_6, v_7, v\}$. This path is chordless since $G$ is a chordal graph.

            \item[$\bullet$] $v_i \notin N_G[v_5]$ and $v_6 \notin
            N_G[v_j]$. In this case, we have a $P_8$ in $G$ induced by the
            vertices $\{v_4, v_2, v_k, v_j, v_5, v_6, v_7, v\}$; thus, $v_kv_5 \in
            E(G)$. From $v_i \notin N_G[v_5]$ we obtain that $v_2 \notin
            N_G[v_5]$ and, thus, $v_6v_k \in E(G)$. Now, $G$ has a
            $3$-sun induced by the vertices $\{v_5, v_k, v_j, v_6, v_i,
            v_3\}$, since we have assumed that $v_iv_5 \notin
            E(G)$, $v_6v_j \notin E(G)$, and other edges do not exist by assumption.
            This is a contradiction to our assumption that $G$ is
            a strongly chordal graph.

            \end{itemize}

        \y Using similar arguments as in Case (B.a)(ii) and Case (C.a), we can prove that if
        $v_3 \notin L$ we come to a contradiction, even in the case where $v_j$ has another
        neighbor $v_x$ in $G_i$ such that $v_xv_5 \notin E(G)$.
        Indeed, in the case where $v_2v_5 \in E(G)$ we can prove
        that $N_G[v_5] \supset N_G[v_j]$ and, thus, $v_3 \in L$. In the case where $v_6v_j \notin
        E(G)$, the vertex $v_x$ will be simple after
        the first iteration during the construction of $\sigma$ and, thus, $v_x < v_y \leq v_i$.

        \item[(D.b)] The vertex $v_3$ has two neighbors $v_5$ and $v_5'$ in $G_y$, such that $v_5v_5' \notin
        E(G)$. Using the same arguments as in Case (B.b), we obtain that in this case $G$ has a $C_4$ which
        is a contradiction to our assumptions.

        \item[(D.c)] The vertex $v_3$ has two neighbors $v_5$ and
        $v_5'$ (where $v_5 \neq v_j$ and $v_5' \neq v_j$) in $G_y$,
        such that $v_5v_5' \in E(G)$, and neither $N_y[v_5] \subseteq N_y[v_5']$
        nor $N_y[v_5'] \subseteq N_y[v_5]$. Using the same arguments as in Cases (B.c) and (C.c), we can prove
        that this case comes to a contradiction.

        \end{itemize}

\end{itemize}


\noindent {\bf Case 2: The vertex $v_i \in I$ and $v_i \notin S$.}
Since $\sigma$ is a strong perfect elimination ordering, each
vertex $v_i \in I$ is simple in $G_i$ and, thus, $\{N_{G_i}[v_k]:
v_k \in N_{G_i}[v_i]\}$ is linearly ordered by inclusion. We will
show that $\{N_G[v_k]: v_k \in N_{G_i}[v_i] \s and \s
\kappa(v_k)=\kappa(v_i) \}$ is linearly ordered by inclusion for
all vertices $v_i \in I$ and $v_i \notin S$. Since $v_i$ is not a
simplicial vertex in $G$, there exist at least two vertices $v_2',
v_j' \in N_G(v_i)$ such that $v_2'v_j' \notin E(G)$. In the case
where there exist no neighbors $v_2'$ and $v_j'$ of $v_i$, such
that $v_2'<v_i<v_j'$ and $v_2'v_j' \notin E(G)$, we have exactly
the same situation as in Case 1, where every neighbor $v_j'$ of
$v_i$ in $G_i$ was joined by an edge with every neighbor $v_2'$ of
$v_i$, such that $v_2'<v_i<v_j'$. Let us now consider the case
where $v_i$ has two neighbors $v_2'$ and $v_j'$, such that
$v_2'<v_i<v_j'$ and $v_2'v_j' \notin E(G)$.

\y Using the same arguments as in Case 1 we can prove that for any
vertex $v_i' \in I$ and $v_i' \notin S$, the set $\{N_G[v_k']:
v_k' \in N_{G_i'}[v_i'] \s and \s \kappa(v_k')=\kappa(v_i') \}$ is
linearly ordered by inclusion. First, we can easily see that for
any two neighbors $v_k'$ and $v_j'$ of $v_i$ in $G_i'$, such that
$v_i'<v_k'<v_j'$ and $\kappa(v_i')=\kappa(v_k')=\kappa(v_j')$, we
can prove that either $N_G[v_k'] \subseteq N_G[v_j']$ or
$N_G[v_k'] \supseteq N_G[v_j']$, by substituting $v_k$ by $v_k'$
and $v_j$ by $v_j'$ in the proof of Case 1. Additionally, we can
see that for any neighbor $v_k'$ of $v_i'$ in $G_i'$, such that
$v_i'<v_k'$ and $\kappa(v_i')=\kappa(v_k')$, we can prove that
either $N_G[v_k'] \subseteq N_G[v_i']$ or $N_G[v_k'] \supseteq
N_G[v_i']$, by substituting $v_k$ by $v_i'$ and $v_j$ by $v_k'$ in
the proof of Case 1. It easy to see that by combining these two
results we obtain that the set $\{N_G[v_k']: v_k' \in
N_{G_i'}[v_i'] \s and \s \kappa(v_k')=\kappa(v_i') \}$ is linearly
ordered by inclusion, for any vertex $v_i' \in I$ and $v_i' \notin
S$.

\y From Cases 1 and 2 we conclude that using the constructed
strong perfect elimination ordering $\sigma$ of $G$, we have
proved that the set $\{N_G[v_k]: v_k \in N_{G_i}[v_i] \s and \s
\kappa(v_k)=\kappa(v_i) \}$ is linearly ordered by inclusion, for
any vertex $v_i \in I$. Thus, the lemma holds.  \s \qed

\yy From Corollary~2.1, we have that $\lambda(G) \geq \alpha(G)$
holds for any graph $G$. Since $\kappa$ is a linear coloring of
$G$ using $\alpha(G)$ colors, it follows that the equality
$\lambda(G)=\alpha(G)$ holds for $G$. Since every induced subgraph
of a strongly chordal graph is strongly chordal \cite{Far83}, we
can construct a strong perfect elimination ordering $\sigma$ as
described above for every induced subgraph $G_A$ of $G$, $\forall
A \subseteq V(G) $; thus, we can assign a coloring $\kappa$ to
$G_A$ with $\alpha(G_A)$ colors. Concluding, the equality
$\lambda(G_A)=\alpha(G_A)$ holds for every induced subgraph $G_A$
of a strongly chordal graph $G$ and, therefore, any strongly
chordal graph $G$ is a linear graph.

\y Therefore, we have proved the following result.

\medskip
\noindent {\bf Lemma 4.2.} {\it Any $P_6$-free strongly chordal
graph is a linear graph.}

\yy From Lemma 4.2, we obtain the following result.

\medskip
\noindent {\bf Lemma 4.3.} {\it If $G$ is a $k$-sun graph ($k \geq
3$), then $G$ is a linear graph.}

\yy \noindent {\sl Proof.} \s Let $G$ be a $k$-sun graph. It is
easy to see that the equality $\alpha(G)=\lambda(G)$ holds for the
$k$-sun $G$. Since a $k$-sun constitutes a minimal forbidden
subgraph for the class of strongly chordal graphs, it follows that
every induced subgraph of a $k$-sun is a strongly chordal graph,
and, thus, from Lemma~4.2 $G$ is a linear graph. \s \qed

\y From Lemmas~4.2 and 4.3, we also derive the following results.

\medskip
\noindent {\bf Proposition 4.5.} {\it Linear graphs form a
superclass of the class of $P_6$-free strongly chordal graphs.}

\yy We have proved that any $P_6$-free chordal graph which is not
a linear graph has a $k$-sun as an induced subgraph; however, the
$k$-sun itself is a linear graph. The interest of these results
lies on the following characterization that we obtain for the
class of linear graphs in terms of forbidden induced subgraphs.

\medskip
\noindent {\bf Theorem 4.2.} {\it Let $\mathcal{F}$ be the family
of all the minimal forbidden induced subgraphs of the class of
linear graphs, and let $F_i$ be a member of $\mathcal{F}$. The
graph $F_i$ is either a $C_n$ ($n \geq 4$), or a $P_6$, or it
properly contains a $k$-sun ($k \geq 3$) as an induced subgraph.}

\section{Concluding Remarks}

In this paper we introduced the linear coloring on graphs and
defined two classes of perfect graphs, which we called co-linear
and linear graphs. An obvious though interesting open question is
whether combinatorial and/or optimization problems can be
efficiently solved on the classes of linear and co-linear graphs.
In addition, it would be interesting to study the relation between
the linear chromatic number and other coloring numbers such as the
harmonious number and the achromatic number on classes of graphs,
and also investigate the computational complexity of the the
harmonious coloring problem and pair-complete coloring problem on
the classes of linear and co-linear graphs.

\y It is worth noting that the harmonious coloring problem is of
unknown computational complexity on co-linear and connected linear
graphs, since it is polynomial on threshold and connected
quasi-threshold graphs and NP-complete on co-chordal, chordal and
disconnected quasi-threshold graphs; note that the NP-completeness
results have been proven on the classes of split and interval
graphs \cite{Asd-Ioan07}. However, the pair-complete coloring
problem is NP-complete on the class of linear graphs, since its
NP-completeness has been proven on quasi-threshold graphs, but it
is polynomially solvable on threshold graphs \cite{Asdre07}, and
of unknown complexity on co-chordal and co-linear graphs.
Moreover, the Hamiltonian path and circuit problems are
NP-complete on the class of linear graphs, since their
NP-completeness has been proven on the class of split strongly
chordal graphs \cite{Mul96}. We point out that, the complexity
status of the path cover problem is open on the class of co-linear
graphs.

\y Finally, it would be interesting to study structural and
recognition properties of linear and co-linear graphs and see
whether they can be characterized by a finite set of forbidden
induced subgraphs.

\end{document}